\documentclass[a4paper,fleqn,usenatbib]{mnras}

\usepackage{newtxtext}
\usepackage{newtxmath}

\usepackage{ae,aecompl}

\usepackage{graphicx}	
\usepackage{amsmath}	
\usepackage{amssymb}	

\newcommand{\Msun}{ h^{-1}{\rm M_{ \odot}}}
\newcommand{\hMpc}{ h^{-1}{\rm Mpc}}
\newcommand{\hkpc}{ h^{-1}{\rm kpc}}
\newcommand{\ihMpc}{ h\,{\rm Mpc}^{-1}}

\DeclareMathOperator\erf{erf}

\title[Modelling of cosmology and baryons]{Modelling the large scale structure of the Universe as a function of cosmology and baryonic physics}

\author[G. Aric\`o et al.]{Giovanni Aric\`o$^{1}$\thanks{E-mail:giovanni\_arico001@ehu.eus (GA)},
Raul E. Angulo$^{1,2}$,
Carlos Hern\'andez-Monteagudo $^{3}$
\newauthor
Sergio Contreras$^{1}$, Matteo Zennaro$^{1}$, Marcos Pellejero-Iba\~nez$^{1}$
\newauthor
\& Yetli Rosas-Guevara$^{1}$
\\
\\
$^{1}$Donostia International Physics Center (DIPC), Paseo Manuel de Lardizabal, 4, 20018, Donostia-San Sebasti\'an, Guipuzkoa, Spain.\\
$^{2}$IKERBASQUE, Basque Foundation for Science, 48013, Bilbao, Spain.\\
$^{3}$Centro de Estudios de F\'isica del Cosmos de Arag\'on, Unidad Asociada CSIC, Plaza San Juan 1, 44001 Teruel, Spain.
}

\date{Accepted XXX. Received YYY; in original form ZZZ}

\pubyear{2019}

\begin{document}
\label{firstpage}
\pagerange{\pageref{firstpage}--\pageref{lastpage}}
\maketitle

\begin{abstract}
We present and test a framework that models the three-dimensional distribution of mass in the Universe as a function of cosmological and astrophysical parameters. Our approach combines two different techniques: a rescaling algorithm that modifies the cosmology of gravity-only $N$-body simulations, and a ``baryonification'' algorithm which mimics the effects of astrophysical processes induced by baryons, such as star formation and AGN feedback. We show how this approach can accurately reproduce the effects of baryons on the matter power spectrum of various state-of-the-art hydrodynamical simulations (EAGLE, Illustris, Illustris-TNG, Horizon-AGN, and OWLS, Cosmo-OWLS and BAHAMAS), to better than 1\% from very large down to small, highly nonlinear, scales ($k\sim5 \, \ihMpc$), and from $z=0$ up to $z\sim2$. We highlight that, thanks to the heavy optimisation of our algorithms, we can obtain these predictions for arbitrary baryonic models and cosmology (including massive neutrinos and dynamical dark energy models) with an almost negligible CPU cost. With these tools in hand we explore the degeneracies between cosmological and astrophysical parameters in the nonlinear mass power spectrum. Our findings suggest that after marginalising over baryonic physics,  cosmological constraints inferred from weak gravitational lensing should be  moderately degraded.
\end{abstract}




\section{Introduction}

Measuring the spatial distribution and growth of mass in the Universe is one of the main probes of the cosmic acceleration and the nature of dark matter \citep[see e.g.][]{Weinberg2013}. Consequently, weak gravitational lensing, which directly maps the cosmic gravitational potential and thus the matter distribution, is among the primary targets of several current and future cosmological surveys (KIDS, DES, HSC SSP, Euclid, LSST).

One of the main advantages of weak lensing is that, since dark matter dominates the mass budget in the Universe, its theory modelling should mostly rely on well-understood physics such as General Relativity. However, although gravitational interactions dominate the nonlinear evolution of mass in the Universe, processes induced by baryonic interactions cannot be ignored. In fact, the accuracy of future measurements will be such that, if neglected, baryonic physics could easily induce a large bias on the cosmological constraints inferred \citep{Semboloni2011,Schneider2019}.

Similarly, various other cosmological observables depend on the distribution of baryons and dark matter on large scales and thus are affected by the same baryonic physics. For instance, thermal and kinetic Sunyaev-Zeldovich effects can probe the cosmological parameters and the law of gravity, but are also sensitive to the distribution and the (thermo-) dynamical state of the gas in and around haloes \citep{McCarthy2014,Hojjati2017,Park2018}. Therefore, in addition to quantifying the impact of e.g. intrinsic alignment of galaxies, non-thermal pressure, or uncertainties in the redshift distribution of background galaxies, the effects of baryons also need to be modelled and understood to great precision in modern cosmology.

Currently, the most accurate way to predict the joint evolution of dark matter and baryons is through cosmological hydrodynamical simulations. These simulations seek to follow the relevant astrophysical processes for galaxy formation along with the nonlinear evolution of the mass field. In general, they predict a suppression of the mass clustering at intermediate scales ($k\sim1~\ihMpc$), and an enhancement on small scales ($k>10~\ihMpc$) with respect to the results from gravity-only (GrO) simulations. The former effect is predominantly due to feedback from AGN and supernovae, whereas the latter to the condensation of baryons into stars \citep[for a review, see][]{Chisari2019}.

Despite a broad agreement among different state-of-the-art simulations, there are discrepancies on the amplitude, redshift evolution, and scales affected by baryonic effects. This is likely a consequence of differences in the numerical scheme, and in the (uncertain but necessary) implementation of various sub-grid recipes \citep{Chisari2018, vandaalen2019}. Furthermore, since most astrophysical processes included in hydrodynamical simulations cannot be predicted ab-initio, they have to be calibrated against observations -- a process which has an intrinsic uncertainty, involve many free parameters, and could moreover depend on the assumed cosmology. All this suggests that we are far from a deterministic modelling of astrophysical processes, and that the impact of baryons is still not understood at a quantitative level. Therefore, the modifications predicted by simulations should not be used at face value in cosmological parameter estimations, and more flexible methods should be seeked.

Different approaches have been adopted to incorporate baryonic effects into the data analysis pipelines. For instance, marginalising over nuisance parameters \citep[e.g][]{HarnoisDeraps2015}; identifying the range of scales potentially affected by baryonic physics and exclude them in data analyses \cite[e.g.][]{DES12018} (but at the expense of discarding a potentially huge amount of cosmological information); or to perform a Principal Components Analysis (PCA) and remove the first components \citep{Eifler2015,Huang2019}.

More general attempts to describe the mass field in the presence of baryons are found in several extensions of the halo-model \citep[e.g.][]{Semboloni2011,Mohammed2014,Fedeli2014,Maed2015,Debackere2019}; in terms of response functions calibrated using Separate Universe simulations \citep{Barreira2019}; in perturbative modelling \citep{Lewandowski2015}; displacing particles according to the expected gas pressure \citep{Dai2018}; or even using machine learning \citep{Troester2019}.

In this work we follow another approach, namely the ``Baryon Correction Model'' (hereafter BCM), initially proposed by \cite{S&T2015} and extended in \cite{S&T2018}. The main idea behind this technique is to split mass elements into 4 categories: galaxies, hot bound gas in haloes, ejected gas and dark matter, whose abundance and spatial distribution are parametrised with physically-motivated recipes. The position of particles in a GrO simulation is then perturbed accordingly.

The advantages of this approach are multiple. Firstly, it is physically motivated and does not rely on any specific hydrodynamical simulation. The approach also captures the nonlinear regime, it takes into account environmental effects, and it provides the three-dimensional matter density field. Finally, it has only a few free parameters which could be constrained directly by observations. Unfortunately, the approach is computationally expensive and relies on the existence of a suite of high-resolution simulations with varying cosmological parameters, both of which limit its usability in real data analyses.

Here, we propose a modified version of the BCM that solves these issues. Our version captures the essence of the original approach but with different assumptions and in a computationally efficient manner. We also extend the model to identify individual simulation particles as part of galaxies, hot gas, cold (ejected) gas, or dark matter. This allows the creation of X-ray, and kinetic and thermal Sunyaev-Zeldovich maps \citep{SZ1970}. Importantly, we also demonstrate that our modified version of the BCM can be accurately combined with the cosmology-scaling algorithm presented in \cite{A&W2010}, so that the BCM can be applied on top of any set of cosmological parameters.

Putting these two ingredients together, we predict the mass power spectrum simultaneously as a function of cosmology and astrophysical parameters. To test the accuracy of our approach, we employ a single GrO simulation with which we reproduce, to better than 1\%, the power spectrum suppression as predicted by various state-of-the-art simulations (EAGLE, Illustris, Illustris-TNG, OWLS, Cosmo-OWLS, BAHAMAS and Horizon-AGN) which adopt different cosmologies and galaxy formation prescriptions. Furthermore, we test the flexibility of our model at $z\le2$, for OWLS, Cosmo-OWLS, BAHAMAS and Horizon-AGN. We find that the accuracy of our model does not degrade at high redshifts, indicating that our assumptions hold over all the broad range of scales and cosmic times considered. As an initial application of the framework developed in this manuscript, we explore the impact of baryons in extracting cosmological information from the mass power spectrum. We stress also the importance of quantifying and correctly propagating the uncertainties of the data model employed, which can be the leading source of error in the forthcoming weak-lensing surveys.

This paper is organised as follows: in \S\ref{sec:sim} we present the $N$-body simulations used in our work, in \S\ref{sec:mbcm} we introduce our baryonic model and quantify its impact in the mass power spectrum; in \S\ref{sec:cosmoscaling} we briefly describe the cosmology rescaling algorithm, its implementation, and its combination to the BCM; in \S\ref{sec:hydro_bcm} we fit state-of-the-art hydrodynamical simulations and provide the best-fitting parameters at $z=0$, studying also their redshift evolution. We explore the cosmological information in the power spectrum in \S\ref{sec:fisher}. We discuss our results and conclude in \S\ref{sec:conclusion}.

\section{Numerical Simulations}
\label{sec:sim}

\subsection{Gravity-Only Simulations}

Our GrO simulations were run with \texttt{L-GADGET-3} \citep{Angulo2012}, an optimised and memory-efficient version of \texttt{GADGET} \citep{Springel2005GADGET}. The initial conditions were generated on-the-fly at $z=49$ using 2nd-order Lagrangian Perturbation
theory and have suppressed cosmic variance thanks to the ``fixed and paired'' technique \citep{Angulo&Pontzen2016}. Gravitational forces were computed using a Tree-PM algorithm with a Plummer-equivalent softening length of $\epsilon_s=5\,\hkpc$. The force and time integration accuracy parameters were chosen so that $z=0$ power spectra are accurate at the $\sim1\%$ level at $k\sim5\,\ihMpc$.

We have built (sub)halo catalogues with a Friends-of-Friends algorithm and a modified version of {\tt SUBFIND} \citep{Springel2001SUBFIND}. The FoF linking length is 2\% of the mean inter-particle separation, $\ell=6.7\,\hkpc$. We kept objects gravitationally bound and resolved with at least 20 particles. Additionally, for all the simulations we stored a set of particles (homogeneously selected in Lagrangian space) diluted by a factor of $4^3$, which we will use as our dark matter catalogue.

We have run a set of three (paired) simulations, with box sides of 64, 128, 256 $\hMpc$ and $192^3$, $384^3$, $768^3$ particles of mass $m_p\approx3.2 \times 10^9~\Msun$. Therefore, Milky-Way like haloes are resolved with $\sim300$  particles. The particle mass was also chosen to achieve a high accuracy on the nonlinear power spectrum \citep{Schneider2016}.

The cosmological parameters were chosen to maximise the accuracy of the rescaling algorithm over a wide range of cosmologies. Specifically: the density of cold dark matter, baryons and dark energy, in units of the critical density, are $\Omega_{\rm cdm}=0.265$, $\Omega_{\rm b}=0.05$, $\Omega_{\Lambda}=0.685$ respectively, the Hubble parameter $H_0=60$ km s$^{-1}$ Mpc$^{-1}$, the spectral index of the primordial power spectrum $n_s=1.01$, the amplitude of the linear fluctuation of the matter density field at $8~\hMpc$, $\sigma_8=0.9$, the optical depth at recombination $\tau=0.0952$, the dark energy equation-of-state parameters assuming a Chevallier-Polarski-Linder parametrisation, \citep{ChevallierPolarski2001,Linder2003}, $w_0=-1$ and $w_a=0$, and the sum of the neutrino masses $\sum m_{\nu}=0$ eV. We refer to \cite{Contreras2020} for further details.

To test the accuracy of the combination of cosmology scaling and baryonic model, we have carried out another paired simulation adopting the Planck 2013 cosmology \citep[][hereafter Planck13]{Planck2013}:
$\Omega_{\rm cdm}=0.2588$, $\Omega_{b}=0.0482$, $\Omega_{\Lambda}=0.6928$, $H_0=67.77$ km s$^{-1}$ Mpc$^{-1}$, $n_s=0.961$, $\sigma_8=0.828$, $\tau=0.0952$, $w_0=-1$, $w_a=0$, $\sum m_{\nu}=0$.
This second simulation has the same number of particles and initial white-noise field as our main simulation, but a box size of $272.4$ $\hMpc$ (instead of $256\,\hMpc$). This volume was chosen to exactly match the box size of our largest simulation after its cosmology was rescaled to Planck13.

All the power spectra shown throughout this paper, unless stated otherwise, are computed by assigning particles in two $512^3$ interlaced grids employing a {\it cloud-in-cell} mass assignment scheme, and using Fast Fourier Transforms. This results into a power spectrum estimation accurate to 1\% up to the grid Nyquist frequency \citep[see][]{Sefusatti2016}. We have checked that using a {\it triangular shaped cloud} scheme our results do not change.

The shot noise contribution is estimated as $1/\bar{n}$, and subtracted. $\bar{n}\,P(k)$ reaches $0.01$ at $k\sim5\,\ihMpc$, a scale $4$ times smaller than the typical wavenumber affected by our choice of softening length ($k\sim\pi (2.7\,\epsilon_s)^{-1}\approx20\,\ihMpc$). Therefore, we will focus on scales  $k\lesssim5\ihMpc$, where we expect numerical noise in our results to be less than 1\%.

\subsection{Hydrodynamical Simulations}

To test the performance of our BCM, we will compare its predictions against measurements from various hydrodynamical simulations. In alphabetical order, these simulations are:

\begin{itemize}
\item {BAHAMAS} \citep{McCarthy2017}: run with \texttt{GADGET3}, calibrated to reproduce the present-day stellar mass function and halo gas mass fractions, with the specific purpose of studying the baryonic impact on the cosmic mass distribution. The simulation we use in this work has a box size of L=$400\,\hMpc$ and it has been run employing a Planck15 cosmology with massive neutrinos \citep{Planck2015} ($\Omega_{\rm cdm}$, $\Omega_{\rm b}$, $\Omega_{\Lambda}$, $A_s$, $h$, $n_s$, $\tau$, $\sum m_{\nu}$)=($0.2589$, $0.0486$, $0.6911$,
$2.116\times10^{-9}$, $0.6774$, $0.9667$, $0.066$, $0.06$).

\item{Cosmo-OWLS} \citep{LeBrun2014}: this set of simulations is an extension of the OWLS simulations, designed to study
cluster-size astrophysics. We use a simulation which include metal-dependent radiative cooling, star formation, stellar and AGN feedback, and have a L=$400 \,\hMpc$ box with a WMAP7 cosmology ($\Omega_{\rm cdm}$, $\Omega_{\rm b}$, $\Omega_{\Lambda}$, $A_s$, $h$, $n_s$, $\tau$)=($0.226$, $0.0455$, $0.72845$, $2.185\times10^{-9}$, $0.704$, $0.967$, $0.085$).

\item{EAGLE} \citep{Schaye2015, Hellwing2016}: a SPH hydrodynamical simulation in a L=$68 \,\hMpc$ box that includes modelling for star formation, thermal AGN feedback, black-hole growth and metal enrichment. The cosmology employed is consistent with Planck13 ($\Omega_{\rm cdm}$, $\Omega_{\rm b}$, $\Omega_{\Lambda}$, $A_s$, $h$, $n_s$, $\tau$)=($0.2588$, $0.0482$, $0.6928$, $2.1492\times10^{-9}$, $0.6777$, $0.9611$, $0.0952$).

\item{Illustris} \citep{Vogelsberger2014}: a $75 \,\hMpc$ box simulated with the adaptive moving mesh code {\tt AREPO} \citep{Springel2010}. Similarly to EAGLE, it includes a wide range of astrophysical recipes, although their implementation and calibration differ. In particular, its thermal AGN feedback has been shown to be over-effective, blowing away most of the baryons inside haloes \citep{vandaalen2019}. The cosmology employed is consistent with WMAP9 ($\Omega_{\rm cdm}$, $\Omega_{\rm b}$, $\Omega_{\Lambda}$, $A_s$, $h$, $n_s$, $\tau$)=($0.227$, $0.0456$,  $0.7274$, $2.175\times10^{-9}$, $0.704$, $0.9631$, $0.081$).

\item{IllustrisTNG-300} \citep{Springel2018}: a $205 \,\hMpc$ box simulated with the same code and an updated version of the physics modelling of {\it Illustris}. Most notably, it features a new kinetic AGN feedback, more in agreement with observations with respect to the previous only-thermal one. The cosmology employed is a Planck15 with massless neutrinos ($\Omega_{\rm cdm}$, $\Omega_{\rm b}$, $\Omega_{\Lambda}$, $A_s$, $h$, $n_s$, $\tau$, $\sum m_{\nu}$)=($0.2589$, $0.0486$, $0.6911$, $2.081\times10^{-9}$, $0.6774$, $0.9667$, $0.066$, $0$).

\item{Horizon-AGN} \citep{Dubois2014}: a $100\,\hMpc$ box run with the Adaptive Mesh Refinement (AMR) algorithm {\tt RAMSES} \citep{Teyssier2002}, which focus on the effects of AGN on various cosmic quantities. The cosmology employed is WMAP7-like ($\Omega_{\rm cdm}$, $\Omega_{\rm b}$, $\Omega_{\Lambda}$, $A_s$, $h$, $n_s$, $\tau$)=($0.226$, $0.0455$, $0.7284$, $1.988\times10^{-9}$, $0.704$, $0.967$, $0.085$).

\item{OWLS} \citep{Schaye2010,vanDaalen2011}:  suite of simulations designed to study different baryonic effects on the cosmic density field. The simulation used in this work has a box of $100\,\hMpc$ and includes AGN feedback. The cosmology employed is from WMAP7  ($\Omega_{\rm cdm}$, $\Omega_{\rm b}$, $\Omega_{\Lambda}$, $A_s$, $h$, $n_s$, $\tau$)=($0.226$, $0.0455$, $0.7284$, $1.988\times10^{-9}$, $0.704$, $0.967$, $0.085$).
\end{itemize}


The power spectra for the mass field together with those for a GrO version of each simulation were kindly provided to us or made publicly available by the authors. To facilitate their comparison, we rebin the original $P(k)$ measurements into the same $k$ bins.

\section{Modified Baryon Correction Model}
\label{sec:mbcm}

 \begin{figure*}
   \includegraphics[width=0.75\linewidth]{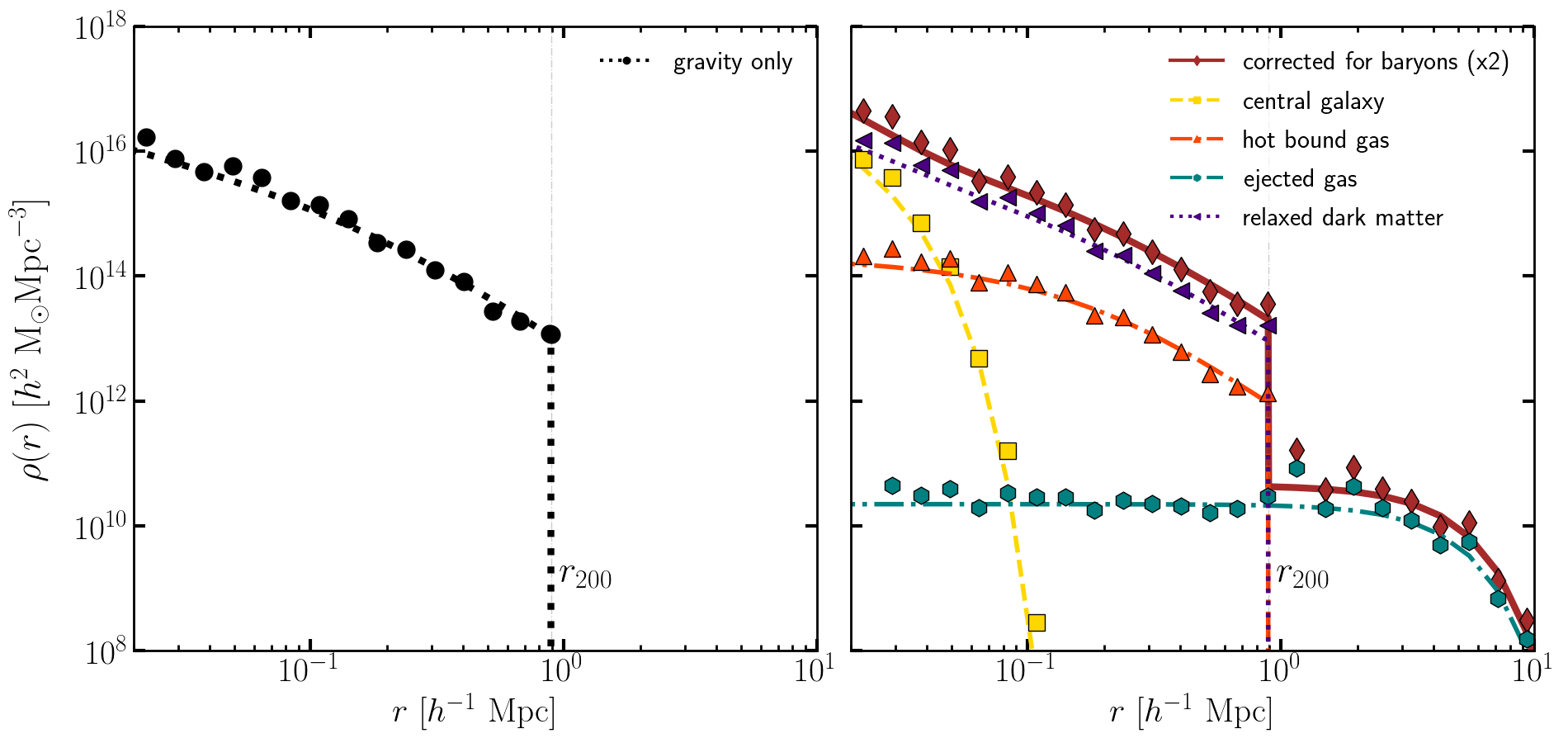}
   \caption{Density profiles of a halo of mass $1.2\times10^{14} ~ h^{-1} ~ M_{\odot}$ and concentration $c\approx4$ at $z=0$. {\it Left Panel:}
   The black circles and dashed line are the measured only gravity profile and its fit, respectively. The initial profile is truncated at $r_{200}$.
  {\it Right Panel:} The brown solid line represents the theoretical total corrected profile, while the diamonds
  the measurements after the displacement of the particles. All the BCM components are displayed according to the legend. Notice how the gas ejected by the AGN feedback is the only component beyond $r_{200}$. The total BCM theoretical and measured density profiles are multiplied by a factor of 2 for display purposes. }
   \label{fig:rho_components}
 \end{figure*}

In this section we describe how we model baryonic processes in a given output of a GrO simulation. Our approach follows closely the Baryonic Correction Model (BCM) proposed by \cite{S&T2015}, with further assumptions to simplify it and speed-up its execution. Most of the recipes of the BCM are given in terms of the host halo mass and radius. Here, we will assume these to be $M_{200}$ and $r_{200}$, the mass and size of a sphere whose average density is equal to 200 times the critical density of the Universe.

\subsection{Overwiew}
The main idea behind the BCM is to capture baryonic effects by explicitly modelling four components: galaxies, hot gas inside haloes,  expelled gas and dark matter:

\begin{itemize}
\item {\it Galaxies} are placed at the minimum of the potential of dark matter (DM) haloes. The mass of galaxies is given by subhalo abundance matching \citep{Behroozi2013}, whereas their internal profile is given by a power law with an exponential cut-off at a scale radius set by observations \citep{Kravtsov2018}.

\item {\it Hot gas} is assumed to be in hydrostatic equilibrium inside DM haloes. The amount of hot gas, $M_{\rm bg}$, is given as a function of the universal baryon fraction, approaching unity for halo masses $M_{\rm h}\gg M_c$, and decreasing as a power law $(M_{\rm h}/M_c)^{\beta}$ for smaller masses, where $M_c$ and $\beta$ are free parameters. The gas profile is described as a power-law with a polytropic index given by the concentration of the host halo, and on scales larger than half virial radius we assume that the gas perfectly traces dark matter.

\item {\it Gas ejected} from its halo is assumed to be distributed isotropically up to a scale $\sim10\,\eta\,r_{200}$. Its density profile is described as a constant with an exponential suppression, consistent with assuming an initial Maxwell-Boltzmann distribution for the velocity of ejected mass particles. The amount of mass ejected is simply given by mass conservation: $M_{\rm ej}=M_{\rm h}-M_{\rm g}-M_{\rm dm}-M_{\rm bg}$.

\item {\it Dark matter} is assumed to be initially described by a Navarro-Frenk-White profile with the same concentration as in the GrO calculation. Posteriorly, the profile is quasi-adiabatically relaxed to account for the modification in the potential induced by the three components described above.
\end{itemize}

\begin{table*}
  \centering
  \begin{tabular}{cc|ccc} 
     \hline
     Parameter & Description & Fiducial Value ($z=0$)\\
     \hline
     $M_c$ & Halo mass scale for retaining half of the total gas &  $3.3\cdot10^{13} \, \Msun$  \\
     $M_1$ & Characteristic halo mass for a galaxy mass fraction $\epsilon=0.023$ & $8.63\cdot10^{11} \, \Msun$ \\
     $\eta$ & Maximum distance of gas ejection in terms of the halo escape radius &  0.54 \\
     $\beta$ & Slope of the gas fraction as a function of halo mass & 0.12\\
     \hline
     \end{tabular}
    \caption{Parameters specifying our model for baryonic physics, and their fiducial values, obtained fitting the BAHAMAS simulation, used throughout this paper. See \S\ref{sec:mbcm} for details on the baryonic model, and \S\ref{sec:hydro_bcm} for details on how the parameters were found.}
  \label{tab:parameters_table}
\end{table*}

The model has four free parameters ($\eta$, $M_c$, $\beta$, $M_1$) with clear physical meaning: $M_c$ is the characteristic halo mass for which half of the gas is retained; $\beta$ is the slope of the hot gas fraction - halo mass relation; $M_1$ is the characteristic halo mass for which the central galaxy has a given mass fraction $\epsilon$ ($\epsilon=0.023$ at $z=0$) and $\eta$ sets the range of distances reached by the AGN feedback. These parameters (all present in the original BCM, even if with slightly different physical meaning and if $M_1$ was fixed) are summarized in Table \ref{tab:parameters_table}, and further details on the whole procedure are given in Appendix \ref{app:profiles}.

For a given halo, we can therefore obtain a prediction for the relative difference for the cumulative mass profile before and after modeling baryons. We then perturb the position of particles inside a halo by applying a displacement $\Phi(r)=r(M_{\rm BC})-r(M_{\rm GrO})$ so that they capture the expected modification induced by baryons.

Finally, we tag each particle in our simulation to be part of one of our four components, and rescale its mass to match the total expected mass in each component. This allows to extend the model to other gas properties e.g. temperature and pressure, and thus to simultaneously model weak lensing and other observables such as X-ray emission or thermal/kinetic Sunyaev-Zeldovich signals. More details of this procedure are provided in Appendix \ref{app:subsampling}.

\subsection{A first example}

To illustrate our model in practice, we have applied it to the haloes of one of the $N$-body simulation described in \S\ref{sec:sim}. In this section we will use the fiducial BCM parameters given in ST15, i.e. $M_c=1.2 \times 10^{14} ~ h^{-1}  ~ {\rm M_{\odot}}$, $\eta=0.5$, $\beta=0.6$, $M_1=2.2 \times 10^{11} ~ h^{-1}  ~ {\rm M_{\odot}}$.

In Fig.~\ref{fig:rho_components} we show a halo of $10^{14} ~ h^{-1} ~ M_{\odot}$ and concentration parameter $\sim4$ at $z=0$. The left panel shows the density profile from the GrO simulation whereas the right panel shows the result after the BCM is applied.
In both panels, the symbols represent the measurements, whereas lines denote the respective analytic descriptions. We can see that the GrO profile is well described by a NFW profile up to its critical radius $r_{200}$. Beyond $r_{200}$, we do not attempt to model the mass distribution, and thus the GrO profile is simply set to zero. On the contrary, the halo density profile  significantly departs from a NFW after  baryons are modelled.

On very small scales, the density essentially follows that of the central galaxy. The hot gas has a NFW slope on large scales but a flatter profile in the inner region. The dark matter is perturbed by the gravitational potential of the other components, resulting in a steeper profile in the inner region and a flatter profile at large radii, albeit the effect is so small that it is not visible by eye.

Beyond the halo boundaries, where the GrO model is null, the density profile is totally constituted by the ejected material. This means in practice that, after the displacement of the particles, the effective density profile in the halo outskirts will be equal to the one given by the GrO simulation plus the ejected component of the BCM.

\begin{figure}
  \includegraphics[width=\linewidth]{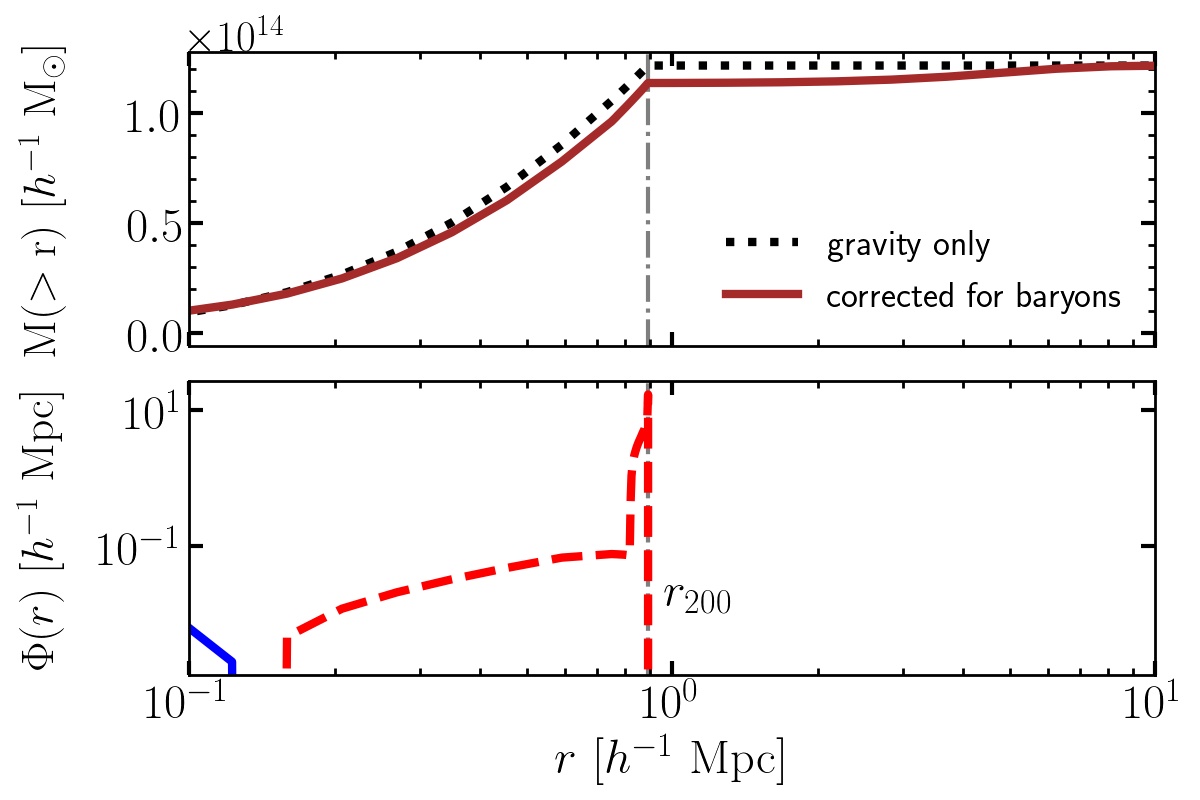}
  \caption{ {\it Upper panel:} Initial gravity-only ($M_{\rm GrO}$, black dotted line) and {\it baryon corrected} ($M_{\rm BC}$, brown solid line) mass profiles. Notice that $M_{\rm GrO}$ is constant after $r_{200}$,       while $M_{\rm BC}$ tends asymptotically to $M_{\rm GrO}$ at large radii, because of the ejected mass. {\it Lower panel:} Displacement field $\Phi(r)=r(M_{\rm BC})-r(M_{\rm GrO})$.
  In radial shells where $M_{\rm GrO}<M_{\rm BC}$ we have that $\Phi<0$ (blue solid line), thus the particles infall toward the centre of the halo. On the contrary, $M_{\rm GrO}>M_{\rm BC}$ implies that $\Phi>0$ (red dashed line) and the particles are pushed away from the centre. Notice also that when approaching $r_{200}$ the displacement becomes of the order of tens of Mpc.}
  \label{fig:mass_displacement_profiles}
\end{figure}

The displacement field and cumulative mass profiles for this halo are shown in Fig.~\ref{fig:mass_displacement_profiles}.
Since we consider initially only the mass within $r_{200}$, the GrO profile is constant for $r > r_{200}$. After modelling baryons, the mass increases up to a scale set by the strength of the AGN feedback. We can also see that the displacement field is largest close to the halo boundary. This implies that these are the mass elements that will be ejected out and describe the expelled gas component.

\begin{figure}
  \includegraphics[width=\linewidth]{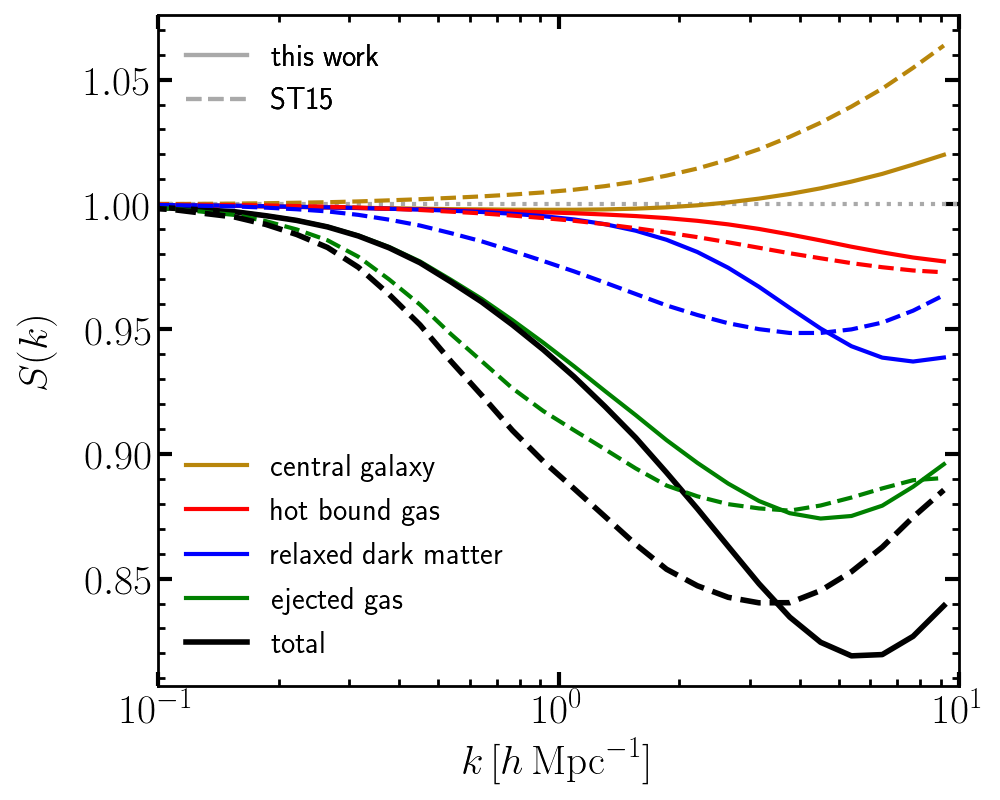}
  \caption{Baryonic effects on the matter power spectrum, defined as $S(k) \equiv P/P_{\rm GrO}$, considering one by one the components of the standard ST15 (dashed lines) and new (solid lines) version the BCM at z=0. The total impact (black) is given by the sum of central galaxy (gold), hot bound gas(red), relaxed dark matter (blue), ejected gas (green) contributions.}
  \label{fig:pk_components_methods}
\end{figure}

The distortions of the halo density profiles translate directly into modifications to the mass power spectrum. In Fig.~\ref{fig:pk_components_methods} we show the ratio of the power spectrum to the GrO one. Coloured lines display the results for the total mass field and for each of the BCM components separately. Consistent with the expectation set by the density profiles, the mass power spectrum is suppressed on intermediate scales owing to the ejected mass, on small scales, the central galaxy counteracts this effect and the power spectrum is enhanced. In \S\ref{subsec:impact} we will investigate systematically these modifications with respect to BCM parameter values.
Fig.~\ref{fig:pk_components_methods} also presents the results for the ST15 method as dashed lines. Although both models agree qualitatively, they disagree in detail. We discuss general and specific differences among them in the next subsection.

\subsubsection{Comparison with Schneider \& Teyssier (2015)}

The main difference of our BCM with respect to that of \cite{S&T2015} is that we assume that baryonic physics acts {\it only} over mass elements within the host halo. Notice that this assumption does not imply a null effect on the large-scale clustering (since the ejected gas does reach large scales), but implies instead that  particles perturbed by baryons were initially inside haloes.
On the contrary, ST15 attempt to model the mass profiles up to infinity,
which in practice means that the displacement tends to zero only at very large distances from the halo centre, and it also implies that in general the displacement of a given mass element receives (a non-commutative) contribution of every single halo in the simulated volume. Furthermore, this approach requires modelling the distribution and clustering of field particles (not belonging to any halo), an operation computationally expensive that cannot anyway take into account halo local environments.

By truncating the profiles at $r_{200}$, thus forcing the displacement of the particles to be zero beyond $r_{200}$, we avoid all these potential issues and remove the non-locality of the model (which appears rather numerical than physical). This also yields a better numerical efficiency as particles inside different haloes can be treated separately, which allows a trivial parallelisation of the algorithm. Our BCM also does not require modelling the mass distribution outside halos, both computationally expensive and uncertain on a halo-by-halo basis. Finally, one could also argue that masses up to the halo virial radius are more correlated to galaxy/gas properties rather than the mass integrated up to infinity.

In our approach we employ similar analytical density profiles and free parameters as those described in \cite{S&T2015}. However, the differences discussed above imply that the BCM parameters affect the nonlinear power spectrum in a somewhat different way.
We now explore the differences in the power spectrum predictions between the ST15 BCM and our version. To do so, we have implemented this BCM following step by step the prescriptions of \cite{S&T2015}, and applied it to our GrO simulation. We compare the power spectra in Fig.~\ref{fig:pk_components_methods} for both models as dashed and solid lines.

For the same model parameters, our implementation predicts less suppression of the power spectrum up to $k\sim3\ihMpc$, and larger suppression on smaller scales. We can understand these discrepancies by examining each BCM component separately (displayed as coloured lines in Fig.~\ref{fig:pk_components_methods}).

On small scales, $k\approx5\,\ihMpc$, the predicted enhancement due to galaxies is smaller than that in ST15 by 2-3 times. Since our halo masses are smaller than in ST15, galaxies are also effectively less massive, which translates into a smaller enhancement of power. We notice that the abundance matching performed by \cite{Behroozi2013} and used in \cite{Kravtsov2018} is calibrated with $M_{200}$ critical, so, unlike ST15, we expect our galaxy mass function to be consistent with observations. Because of the lower halo mass, haloes also have less gas, both ejected and in equilibrium. Therefore, we expect a weaker impact of gas components on the matter power spectrum, which is indeed what is displayed by blue and green lines. Finally, in the ST15 implementation the dark matter quasi-adiabatic relaxation causes a suppression of $\approx5$ \%, affecting large scales, whereas in ours the effect is negligible at $k<2~\ihMpc$. This is also expected by the weaker modification of the gravitational potential caused by the baryons, combined with the assumption that the halo back reaction is negligible at scales larger than $r_{200}$.

\subsection{Numerical implementation}

The concentration of each of these haloes is found by fitting a NFW form to the mass profile computed over 20 bins uniformly spaced in $\log(r/r_{200})$ over the range $[3\, \epsilon_s /r_{200}, 1]$.
We assume that we can correctly fit the density profiles of haloes with more
than 500 particles, and we compute the baryonic corrections for haloes with more than 10 subsampled particles, i.e. $M_{\rm h}\ge2\cdot10^{12} \, \Msun$. We argue that this mass limit is suitable for our analysis, provided that the dominant contribution of the large-scale baryonic effects is given by haloes of $M_{\rm h}\ge\cdot10^{13} \, \Msun$, as we show in Appendix \ref{app:convergence}. \\
To increase the computational efficiency of the BCM, the density profiles are computed and stored directly on-the-fly by our $N$-body code. Furthermore, at each output, particles are sorted according to halo membership, and their relative distances to the parent halo centre stored. Additionally, the concentrations of all the haloes are computed and stored in post-processing for each snapshot of the simulation. All this allows to be able to quickly apply the BCM exploiting OpenMP and MPI parallelisation. On average, the whole procedure takes $\approx5$ ($\approx0.5$) seconds on 4 threads when applied to our biggest (smallest) simulation. \\
The BCM displacement field is found by inverting the mass profiles, and in this case the truncation produces a thin shell of very large displacement at radii approaching $r_{200}$. The shape of this thin shell affects the matter density field at all the scales larger than $r_{200}$, thus we refine the radial bins in that region to have more precision in the matter distribution on large scales.

\subsection{Impact of baryons on the power spectrum}
\label{subsec:impact}

In this subsection we study the range of possible distortions of the matter power spectrum allowed by the BCM.

\begin{figure*}
  \includegraphics[width=0.7\linewidth]{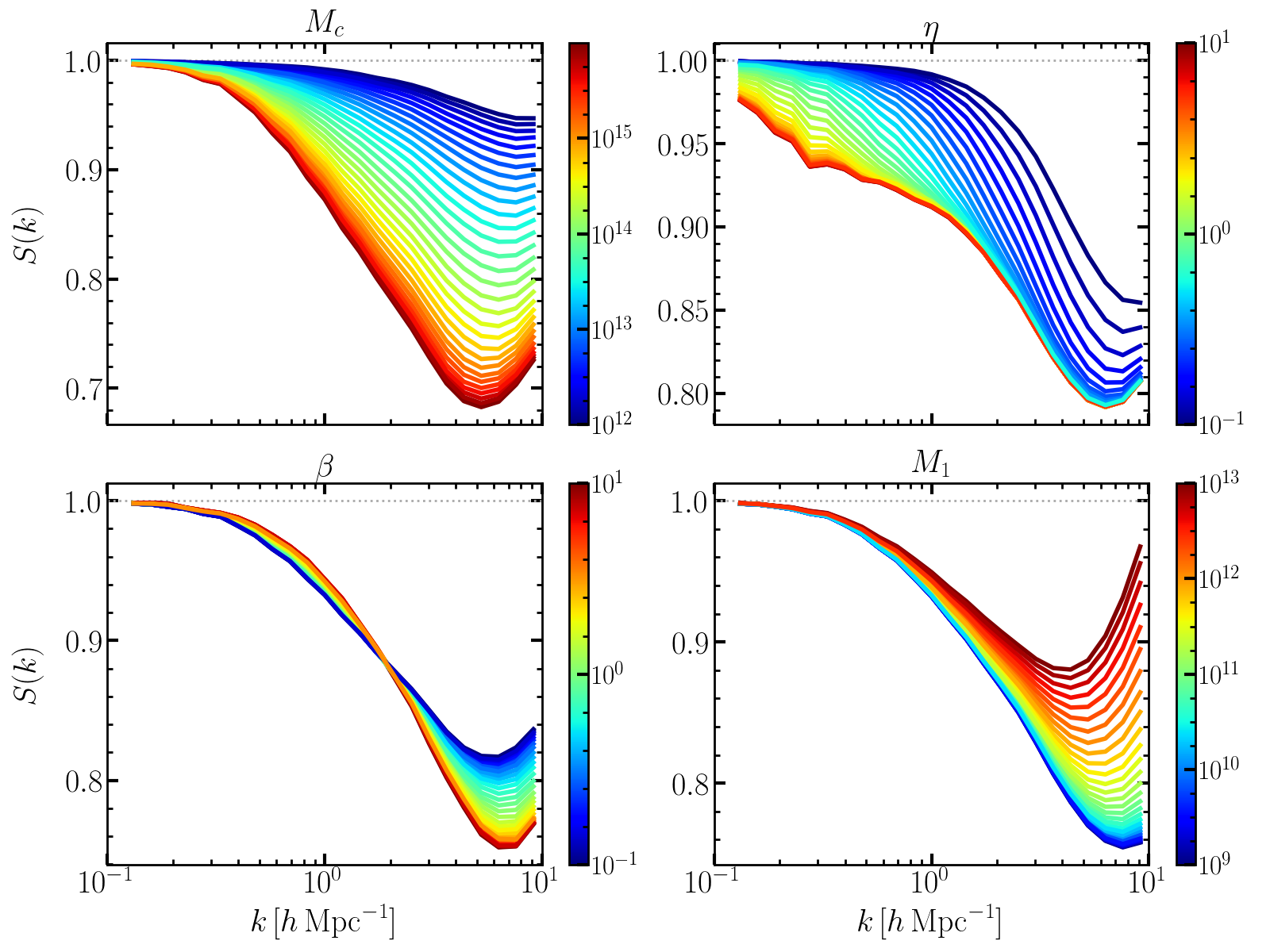}
  \caption{Modifications to the matter power spectrum at $z=0$ caused by baryons, $S(k) \equiv P/P_{\rm GrO}$. Each panel varies one of the four free parameters of the baryon correction model ($M_c$, $\eta$, $\beta$, $M_1$) while keeping the other three fixed at their fiducial value.}
  \label{fig:bcm_pk_dependence}
\end{figure*}

In Fig.~\ref{fig:bcm_pk_dependence} we display the mass power spectra obtained after applying the BCM to our fiducial GrO $N$-body simulation. Each panel varies a single parameter of the model while keeping the other three fixed. Bluer (redder) colors represent low (high) parameter values.

The top left panel varies $M_c$, the typical mass of haloes that have lost half of their gas, in a logarithmic range $[12,16] \, \Msun$. For low values of $M_c$, the power spectrum barely changes owing to the relatively minor contribution that $\lesssim10^{13}\,\Msun$ haloes have to the mass power spectrum. As $M_c$ increases, however, more haloes lose baryons and the power spectrum is suppressed more.

The larger the halo mass, the larger the scale over which baryons are redistributed by feedback, thus the power spectrum suppression affects progressively larger scales. Eventually, when $M_c\sim10^{15}\,\Msun$ the abundance of haloes drops and the power spectrum converges.

How rapid the baryon fraction decreases with halo mass is controlled by $\beta$, which is varied in the logarithmic range $[-1,1]$ (bottom left panel). We can see that the impact of this parameter is smaller if compared to that of $M_c$. Higher (lower) values produce a faster (slower) transition to haloes devoid of gas, and consequently the power spectrum is tilted, being  more (less) suppressed on small scales and less (more) on large scales.

The top right panel varies $\eta$ (in the same range as $\beta$) and consequently the radius up to which the ejected gas will settle in. The larger the value of $\eta$, the further the gas is expelled and therefore the larger the scales that are suppressed. In principle there is no bound on the minimum wavenumber affected, in fact, in the limit of $\eta\rightarrow\infty$, all wavelengths are affected. On the other hand, as $\eta$ decreases, the expelled gas remains very close to its initial position and the power spectrum barely changes.

Finally, the bottom right panel varies $M_1$, the typical mass of haloes with a central galaxy of $0.023 \, M_1 \, \Msun$, in the logarithmic range $[9,13]\, \Msun$. There are two separate trends visible in this plot. Firstly, as we increase $M_1$ a larger fraction of baryons is transformed into stars, which in turn reduces the amount of expelled gas and consequently, the power spectrum suppression is reduced up to $k\approx3\,\ihMpc$. On smaller scales, the contribution of stars in the modelled galaxies becomes important, which increases the amplitude of these Fourier modes.

Overall, we see that the BCM has flexibility to model many different physical scenarios, but, at the same time, not every possible $P(k)$  modification is allowed. In fact, the modifications are constrained to certain regions and have very specific dependences with the wavelength. Therefore, it is not guaranteed that the model is able to accurately reproduce the predictions of state-of-the art hydrodynamical simulations. We explore this in \S\ref{sec:hydro_bcm}.

\section{Cosmology scaling of gravity-only simulations}
\label{sec:cosmoscaling}

\begin{figure}
  \includegraphics[width=\linewidth]{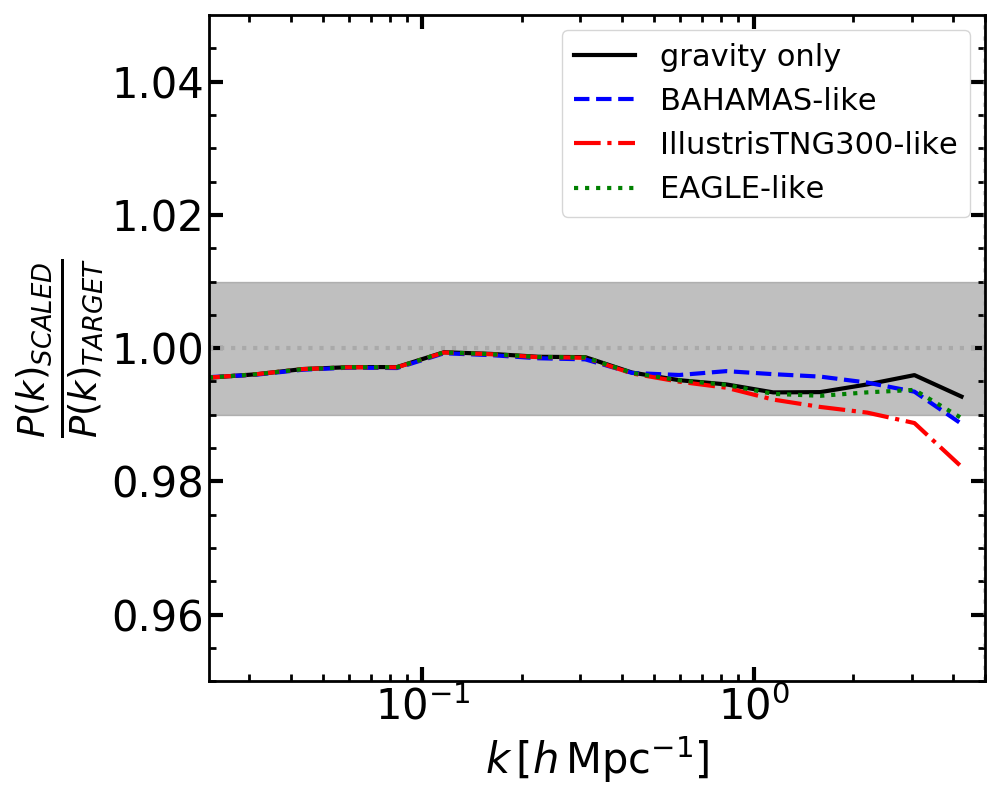}
  \caption{Ratio of the mass power spectra at $z=0$ of two simulations in the cosmology preferred by Planck13: one scaled from our fiducial cosmology, $P_{\rm scaled}(k)$, and the other carried out directly with Planck13, $P_{\rm target}(k)$. Black lines display results of GrO simulations, whereas coloured lines do so for simulations where baryons are explicitly modelled in the BCM with parameters mimicking EAGLE, Illustris-TNG, and BAHAMAS, as indicated by the legend. The grey band marks a discrepancy of 1\%.}
  \label{fig:pk_scaling}
\end{figure}

The BCM enables a flexible modelling of baryonic effects provided a suite of high-resolution GrO simulations with varying cosmological parameters. Here we will show that these GrO predictions can be obtained accurately and efficiently using cosmology-rescaling techniques.

The main idea of a cosmology rescaling is to transform the length, time, and mass units of the outputs of a $N$-body simulation, so that it predicts the nonlinear structure expected in arbitrary-many nearby cosmologies \citep{A&W2010,AnguloHilbert2015}. The algorithm has been extensively tested \citep[][]{Ruiz2011,Renneby2018,Mead2014a,Mead2014b,Mead2015,Zennaro2019,Contreras2020}, and has recently been extended to cover massive neutrino cosmologies \citep{Zennaro2019}, where the redshift and scale dependence of the growth factor induced by the neutrinos is computed with the public code {\tt reps} \citep{Zennaro2017}.

Here we employ the latest incarnation of the cosmology-rescaling, which, in addition to the units transformation, includes a correction of large-scale modes using 2nd order Lagrangian Perturbation Theory and a correction of small-scale modes. For the latter, the algorithm displaces the particles inside haloes to account for the cosmology-dependence of the concentration-mass-redshift relation. For further details we refer the reader to \citep{Contreras2020}.

In order to maximise the accuracy of the method, we should have a snapshot taken exactly at the transformed cosmic time. In general, if we rescale a pre-existing simulation, we can apply only the time transformations allowed by the finite number of snapshots stored, decreasing the accuracy of the method. Obviously, the more snapshots stored the higher the accuracy achieved. We have stored 94 snapshots on the expansion factor interval $a=[0.02,1.25]$ (notice that the simulation is run ``to the future'', $z<0$, making possible the scaling to extreme cosmologies). To increase even more the accuracy of the method, we apply the algorithm to the two snapshots closest to the scaling target time, interpolating afterwards the chosen summary statistics. Having two snapshots taken at cosmic expansion factors $a_0$ and $a_1$, and scaled expansion factor $a_{*}$ such that $a_0<a_*<a_1$, the interpolated scaled power spectrum reads

\begin{equation}
P(a_*) = P(a_0)\cdot \left (1 - \frac{a_*-a_0}{a_1-a_0} \right) + P(a_1) \cdot \left( \frac{a_*-a_0}{a_1-a_0} \right),
\label{eq:interpolation}
\end{equation}

\noindent where $P(a_0)$ and $P(a_1)$ are the power spectra measured rescaling the two snapshots at $a_0$ and $a_1$,
respectively.

\subsubsection{Scaling of the halo catalogue}

The halo catalogue is directly rescaled, to avoid to run {\tt SUBFIND} on the rescaled distribution of particles. Within the standard scaling algorithm, the density profiles should be simply $\rho_*(r)=m_*/s_*^3\rho(r)$, where $m_*=s_* ^3 \Omega_{\rm m,T}/\Omega_{\rm m,O}$ is the mass scale factor, $s_*$ is the length scale factor and the ``T'' and ``O'' subscripts refer to the target and original cosmologies, respectively. Accordingly, the NFW parameters should be rescaled as $r_{s,*}=s_*r_{s}$ and $\rho_{c,*}=m_*/s_* ^3 \rho_c$. The concentration correction adds an extra displacement which we take into account as $\rho_*^{\dagger}(r)=\rho_*(r)+\Delta \rho(r)$, where $\Delta \rho(r)$ is the difference between the NFW profiles of two haloes with the concentration computed within target and scaled cosmology. The scale radius of the NFW is then $r_{s,*}^{\dagger}=r_{s,*}+\Delta r_s$ and the characteristic density $\rho_{c,*}^{\dagger}=\rho_{c,*}+\Delta \rho_c$. The critical radius and mass $r_{200}$ and $M_{200}$ are then found with a minimisation over the new NFW halo profile and within the target cosmology.

\subsubsection{Joint performance of cosmology scaling and BCM}

The scaling algorithm provides highly accurate predictions for the mass power spectrum -- better than $3\%$ at $z\le1$ over the whole range of $\Lambda$CDM-based cosmologies currently viable, and over a wide range of scales $0.01 -  5\,\ihMpc$ \citep{Contreras2020}. Similarly, the halo mass function is reproduced with an accuracy better than 10\% \citep{A&W2010}. In the following we will show that the algorithm also provides high-quality predictions for the mass clustering when used along with the BCM.

To quantify the accuracy of the method, we have rescaled our fiducial simulation (c.f. \S\ref{sec:sim}) to the cosmological parameters preferred by Planck13. We then apply the BCM to the rescaled output and compare to the results obtained by applying the BCM to a simulation directly carried out with a Planck13 cosmology.

Fig.~\ref{fig:pk_scaling} shows the ratio of the power spectra at $z=0$ for three different sets of BCM values. These sets were chosen so that they accurately describe the baryonic effects in the EAGLE, Illustris-TNG, and BAHAMAS simulations. For comparison, we also show the precision when just rescaling GrO outputs. For all models considered the accuracy of the cosmology rescaling is preserved at a very high level, adding no more than $1\%$ additional uncertainty over the rescaling of GrO simulations. Notice that we find similar results for Illustris, Horizon-AGN, OWLS and Cosmo-OWLS, even if not shown in figure for display purposes.

\section{Fitting the state-of-the-art hydrodynamical simulations}
\label{sec:hydro_bcm}

\begin{figure*}
\centering
  \includegraphics[width=0.7\linewidth]{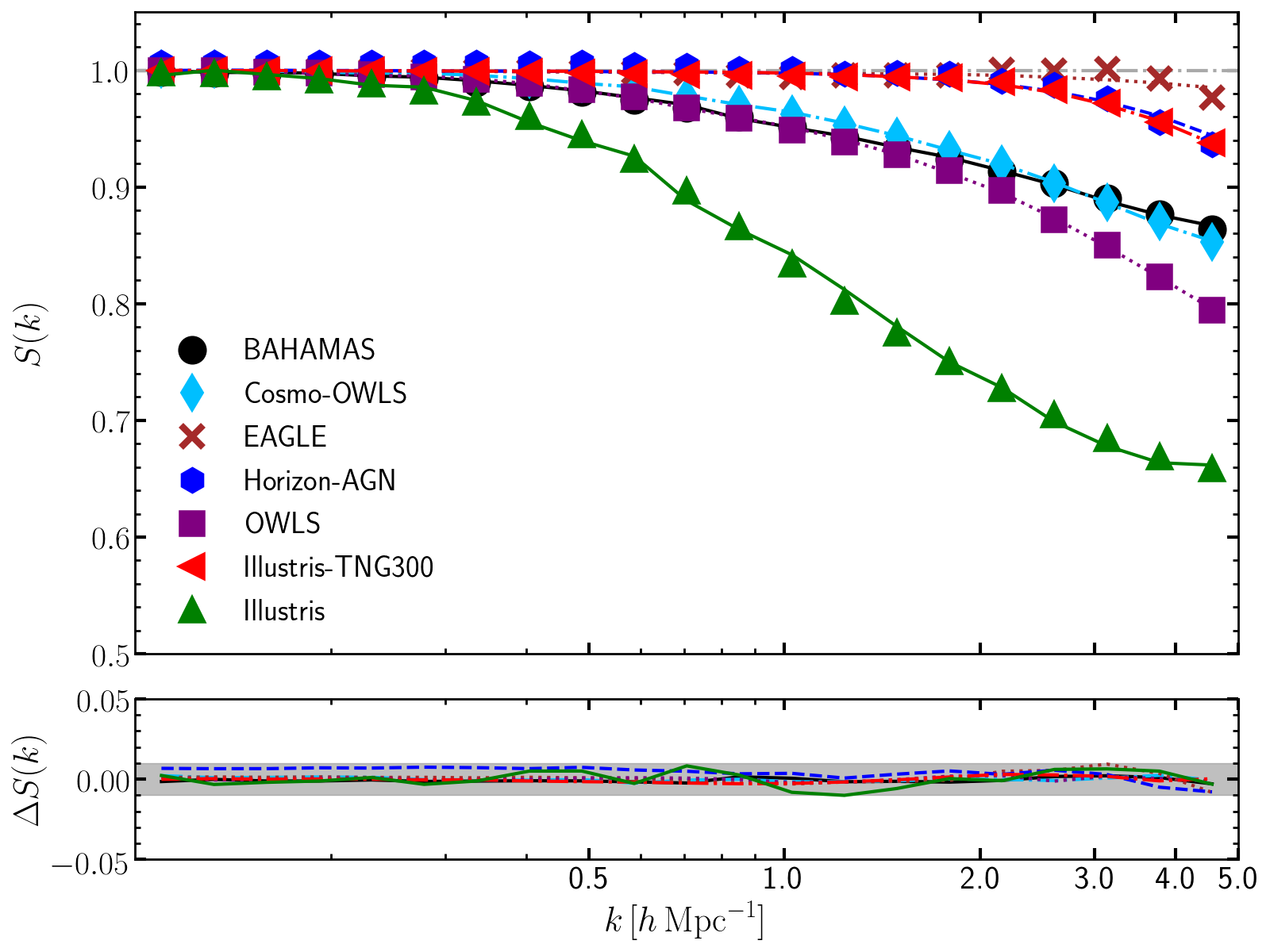}
  \caption{ {\it Upper Panel:} Measurements of the baryonic impact to the matter power spectrum, $S(k) \equiv P/P_{\rm GrO}$, in different hydrodynamical simulations according to the legend (symbols), compared against our respective best-fitting (solid lines). {\it Lower Panel:} Difference between measurements and best-fitting. The grey shaded band marks the 1\% difference.}
  \label{fig:hydro_models}
\end{figure*}

In this section we will explore if the BCM is able to correctly describe the baryonic effects predicted in seven different state-of-the-art hydrodynamical simulations: EAGLE, Illustris, Illustris-TNG, Horizon-AGN, OWLS, Cosmo-OWLS, and BAHAMAS. These simulations adopt different cosmological parameters, sub-grid physics, and values for the free parameters (owing to different strategies and observations used to calibrate them). Therefore, this exercise will test our ability to simultaneously model cosmology and astrophysics.

\subsection{Simulation data \& BCM parameter sampling}
\label{subsec:bcm_pars}

For each of the seven hydrodynamical simulations, we fit the ratio of the mass power spectrum with respect to its GrO counterpart: $S(k)=P_{\rm hydro}/P_{\rm GrO}$.
We interpolate $S(k)$ in 20 data points uniformly spaced in log-$k$ over the range $[0.1-5]$ $\ihMpc$, to have consistent measurements for all the simulations.
We use an empirical approach to estimate the covariance of $S(k)$. First, we assume:

\begin{equation}
\mathcal{C}_{S,ij} = \mathcal{E}(k_i) \mathcal{K}(k_j,k_i) \mathcal{E}_j^{T}(k_j),
\label{eq:suppression_covariance}
\end{equation}

\noindent where $\mathcal{E}$ is an envelope function that describes the typical amplitude of the uncertainty as a function of wavenumber, and $\mathcal{K}(k)$ the correlation of this uncertainty, which we model as a Gaussian distributed random variable $\mathcal{K}=\mathcal{N}(|k_i-k_j|, \ell)$.

We set the magnitude of each term based on the intra-data variance as a function of scale. Specifically, on large scales we assume $\mathcal{E}$ to be constant, with a correlation length $\ell=0.1\,\ihMpc$.
To model the small-scale noise we use $\mathcal{E}=[1 + 0.5\,\erf(k-2)]fS(k)$, where $f=0.6\%$ for BAHAMAS, Cosmo-OWLS, EAGLE and Illustris-TNG300, $f=0.8\%$ for OWLS and $f=2\%$ for Illustris, with a longer correlation length $\ell=0.5\,\ihMpc$.

We should take particular care in the case of Horizon-AGN. The snapshots of the hydrodynamical run were taken at slightly different redshifts with respect to the GrO. We correct at first order this effect by rescaling the power spectra normalised by the growth factors at the correct expansion factors according to Eq.~A2 of \cite{Chisari2018}. However, even after this correction there is still a 1\% disagreement on large scales, arguably given by a difference in the number of particle species with which the two simulations have been carried on \citep{Angulo2013, Chisari2018, vandaalen2019}. For this reason, we set the amplitude of the envelope functions for Horizon-AGN to 1\%.

To fit the ratio measurements, $S(k)$, we first rescale our fiducial GrO simulation to match the cosmology of each of the seven simulations. We then compute the power spectra before and after applying our BCM. For this procedure we use a $64 \, \hMpc$ simulation, avoiding to use its paired and the interpolation between snapshots described in \S\ref{sec:cosmoscaling}. We show in Appendix \ref{app:convergence} that this choice will not affect the final results, since we expect the suppression $S(k)$ to be converged at 1\% level. We have furthermore tested that the small differences in redshift between target and rescaled power spectra is at first order canceled out in the ratio.

We recall that the BCM is fully specified by 4 parameters: $\vartheta=(M_1, M_c, \eta, \beta)$. The prior for these parameters are assumed to be flat in log space over the range: $\log M_1 \in [9,13] \, \Msun$, $\log M_c \in [12,16] \, \Msun$, $\log \eta \in [-1,1]$, $\log \beta \in [-1,1]$. We note that with this prior choice, the ejected radius of each halo is defined such that $r_{ej} \ge r_{200}$.

We assume that the probability of measuring $S(k)$ is given by a multivariate normal distribution with the covariance matrix provided in Eq.~\ref{eq:suppression_covariance}. We define our likelihood as

\begin{equation}
    \mathcal{L(\vartheta | \mathcal{D})} \propto \exp \left[ -\frac{1}{2} \sum_{k} \left( \frac{S(k)_{\mathcal{D}}-S(k)_{\rm \vartheta}}{\Sigma_{S(k)}} \right)^2 \right],
\label{eq:likelihood}
\end{equation}

\noindent where the subscripts $\mathcal{D}$ and $\vartheta$ refer to data and theoretical model, respectively, and $\Sigma_{S(k)}$ is the diagonal of $\mathcal{C}_{S,ij}$ defined in Eq.\ref{eq:suppression_covariance}. We sample the posterior probability with the affine invariant MCMC algorithm {\tt emcee} \citep{Foreman2013}, employing 8 walkers initialised with a latin-hypercube to optimise the hyper-volume spanned. Each walker has 2500 steps with a burn-in phase of 1000, for a total number of 12000 sampling points excluding the burn-in.
We highlight that thanks to the heavy optimizations of all the codes involved, a chain step can be carried out in less than 6 seconds on a common laptop.

\begin{figure}
\centering
  \includegraphics[width=\linewidth]{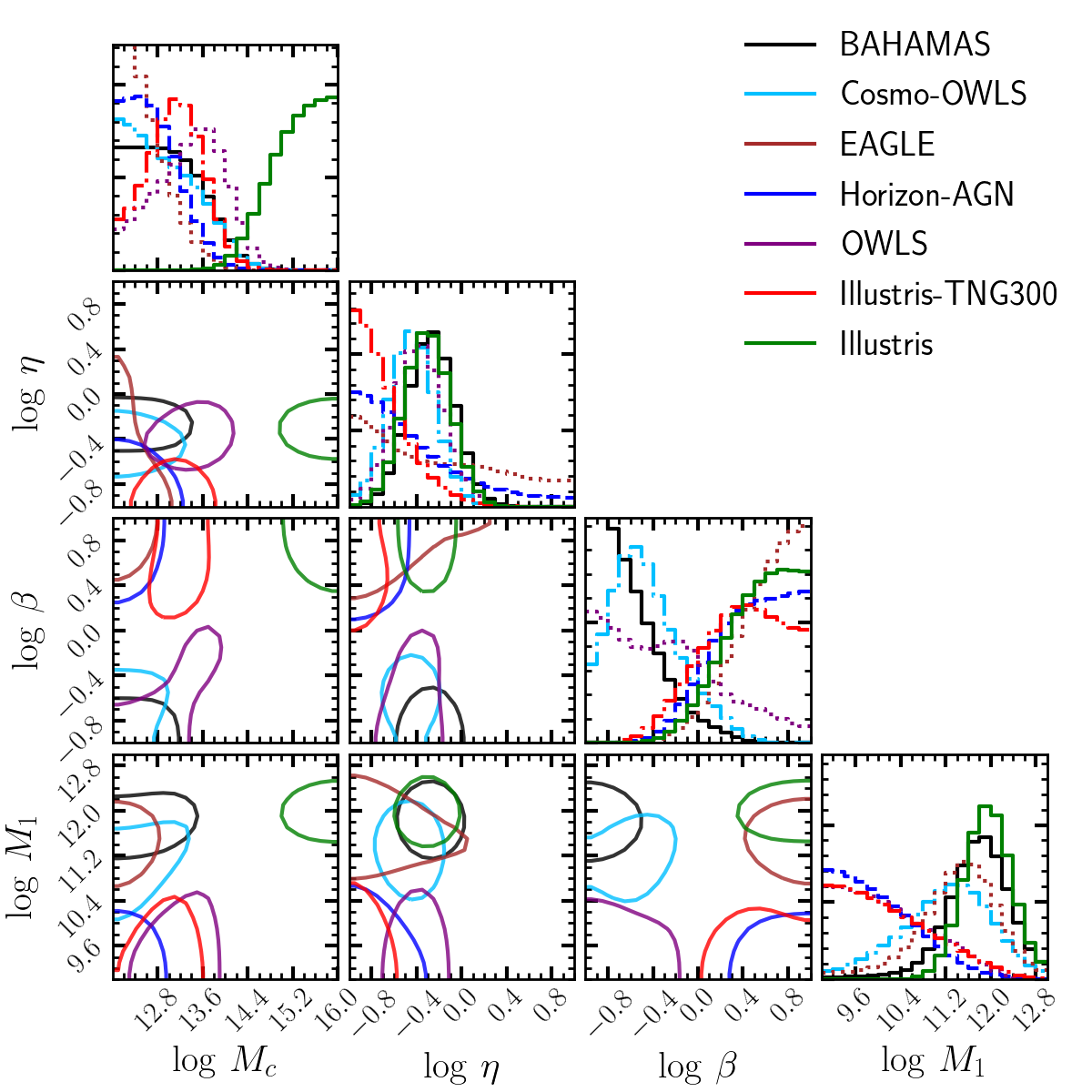}
  \caption{1$\sigma$ credibility levels of the free parameters of our Baryonic Correction Model, obtained by fitting the suppression in the power spectrum at $z=0$  for EAGLE (brown), Horizon-AGN (blue), Illustris (green), Illustris-TNG (red), OWLS (purple), Cosmo-OWLS (light-blue) and BAHAMAS (black). The upper subplots show the marginalised posterior PDF of the baryonic parameters. The best-fitting models are shown in Fig.~\ref{fig:hydro_models}. }
  \label{fig:hydro_contours}
\end{figure}

\subsection{best-fitting parameters}

We now present the best fits and constraints on the BCM parameters as estimated from various hydrodynamical simulations.

Fig.~\ref{fig:hydro_models} compares the measured suppression $S(k)$ with that predicted by our method evaluated with the best-fitting parameters. Remarkably, we can see that the BCM is an excellent description of the data at all scales considered. This is quantified in the bottom panel, which displays the difference between the data and the best fit model, thus can be interpreted as the fractional accuracy of the model in predicting the full power spectrum. For all simulations and scales, this is better than 1\%.

In Fig.~\ref{fig:hydro_contours} we show the $1\sigma$ credibility levels and the marginalised posterior probability density functions (PDFs) of the BCM parameters. The best-fitting values, together with the means and modes of the marginalised posteriors, are provided in Table \ref{tab:simulation_table}.

We can see that there is a broad agreement between the preferred values for some parameters and between a subset of simulations, however, in general different hydrodynamical simulations lie on different regions of the BCM parameter space, as a consequence of the very different predictions for the suppression $S(k)$ owing to the differences in their physics implementation.

The Illustris simulation \citep{Vogelsberger2014} displays the largest suppression, $\approx$ 35 \% at $k\approx6\,\ihMpc$, whereas the EAGLE run presents the weakest, 2\% on the same scale. Other simulations fall in between, with Illustris-TNG and Horizon-AGN providing almost identical suppressions, as well as Cosmo-OWLS and BAHAMAS, at least on the scales considered.

Consistent with this picture, the expected value of $M_c$ is the largest for Illustris and the smallest for EAGLE: $\approx10^{15}$ and $\approx10^{12}\,\ihMpc$, respectively, with the other simulations in between.
Interestingly, Illustris, BAHAMAS, Cosmo-OWLS and OWLS prefer roughly the same value of $\eta\approx0.5$, whereas Illustris-TNG300, Horizon-AGN and EAGLE are consistent with each other and prefer much smaller values, $\eta\approx0.15$, consistent with almost no ejected gas to large distances. BAHAMAS, Cosmo-OWLS and OWLS prefer small $\beta$ values, $\beta\lesssim0.5$, in contrast with the other simulations, which have rather larger values, $\beta\gtrsim2.5$. The expected values of $M_1$ for Horizon-AGN, Illustris-TNG and OWLS are $\lesssim10^{10}\,\Msun$, whereas for all the others $M_1\gtrsim10^{11}\,\Msun$.

Finally, we note that there are rather weak degeneracies among parameters, which supports the idea that the BCM is a general and minimal modelling of baryonic effects in simulations. It is also clear that there is no consensus on the magnitude of baryonic corrections, and thus the need for a flexible modelling for cosmological data analysis.

\begin{table*}
  \centering
  \begin{tabular}{c|cccc} 
     \hline
     Simulation & $M_c \, [10^{14} \, \Msun]$ & $\eta$ & $\beta$  &   $M_1 \, [10^{11} \, \Msun]$\\
     \hline

     BAHAMAS &  (0.38, 0.33, 0.08) &  (0.53, 0.54, 0.53) &  (0.47, 0.12, 0.22) &  (10.85, 8.63, 5.63) \\
     Cosmo-OWLS &  (0.04, 0.01, 0.07) &  (0.35, 0.35, 0.36) &  (0.25, 0.22, 0.34) &  (1.61, 2.09, 1.54) \\
     OWLS &  (0.4, 0.45, 0.24) &  (0.46, 0.43, 0.41) &  (0.67, 0.8, 0.45) &  (0.01, 0.04, 0.14) \\
     Horizon-AGN &  (0.12, 0.05, 0.04) &  (0.15, 0.17, 0.35) &  (6.38, 8.31, 3.15) &  (0.07, 0.02, 0.09) \\
     Illustris-TNG300 &  (0.23, 0.19, 0.12) &  (0.14, 0.15, 0.18) &  (4.09, 2.56, 2.56) &  (0.22, 0.03, 0.14) \\
     Illustris &  (66.48, 91.03, 22.1) &  (0.49, 0.5, 0.5) &  (6.36, 5.42, 3.66) &  (9.44, 9.65, 8.85) \\
     EAGLE &  (0.18, 0.01, 0.03) &  (0.14, 0.11, 0.58) &  (9.65, 6.23, 4.23) &  (11.15, 4.2, 2.52) \\

     \hline
     \end{tabular}
    \caption{For each BCM parameter we tabulate the best-fitting, the mode and the mean values of the marginalised posterior PDF. See \S\ref{sec:mbcm} for details on the baryonic model, and \S\ref{sec:hydro_bcm} for details on how the values were found.}
  \label{tab:simulation_table}
\end{table*}

\begin{figure}
\centering
  \includegraphics[width=\linewidth]{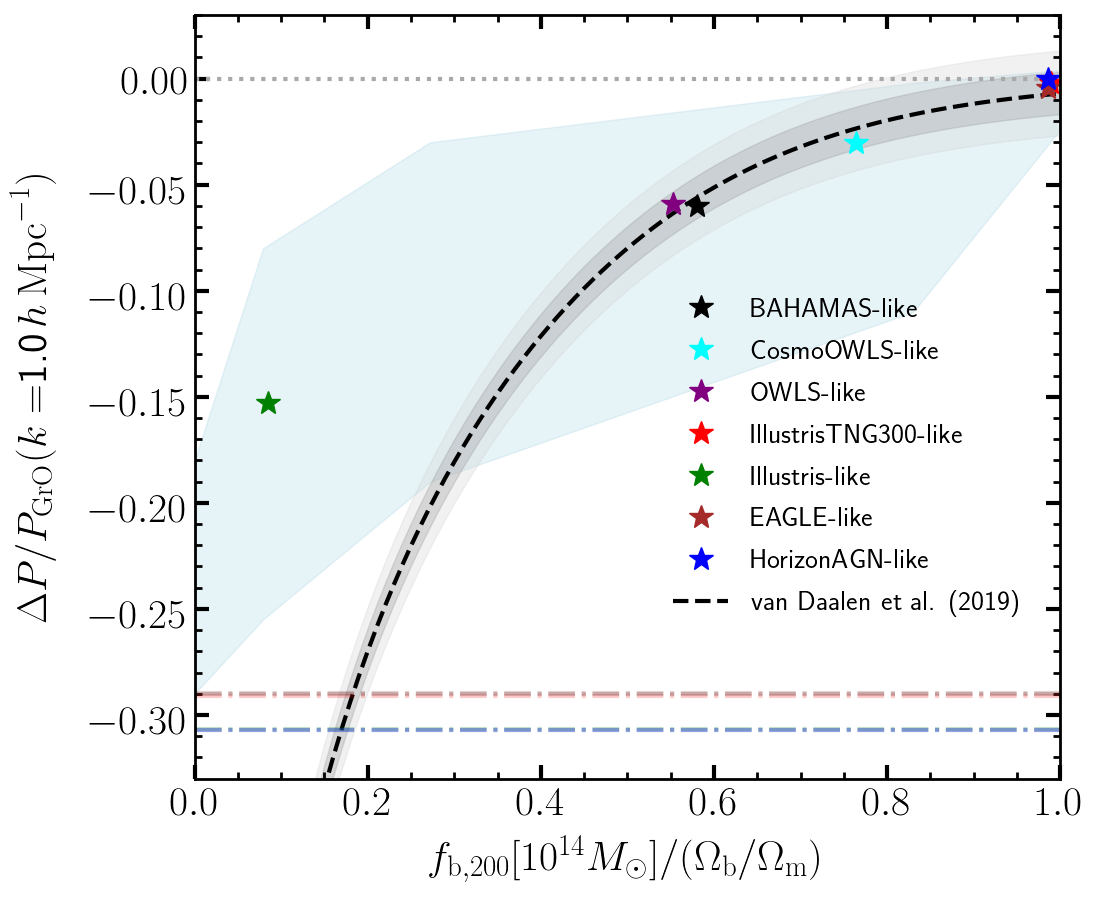}
  \caption{Baryonic impact on the matter power spectrum at $k=1\, \ihMpc$, defined as $\Delta P(k) / P(k)$ as a function of the halo baryon fraction for haloes of $10^{14} \, {\rm M}_{\odot}$.
   The star symbols correspond to the quantity measured in our simulation using a feedback model that resemble the hydrodynamical simulations specified in the legend, in halo mass interval of [$6\times10^{13}$,$2\times10^{14}$] M$_{\odot}$. The light blue shaded area marks the region allowed by the BCM, varying the parameters
    within the priors in logarithmic space $\log M_1 \in [9,13]$, $\log M_c \in [12,16]$, $\log \eta \in [-1,1]$, $\log \beta \in [-1,1]$. The dashed-dotted lines represents the maximum theoretical suppression given by
   $(1-\Omega_{b}/\Omega_{m})^2-1$, the different colors being referred to the simulation cosmology according to the legend. The black dashed line is the fit provided by \protect\cite{vandaalen2019}, being the grey and light grey shaded areas the 1\% and 2\% deviations, respectively.}
  \label{fig:feedback}
\end{figure}

\subsection{Relation to the baryon fraction in clusters}

Although hydrodynamical simulations are calibrated to reproduce several observables properties, they make specific choices for various processes of their sub-grid physics.

Recently, \cite{vandaalen2019} analysed a suite of simulations from the BAHAMAS, OWLS, and Cosmo-OWLS projects to study how the initial mass function, supernovae and AGN feedback, and metal enrichment recipes impact the power spectrum. Regardless of these choices, they found a tight correlation ($\approx$1\%) between the mean baryon fraction inside haloes and the power spectrum suppression. These relations also held for the EAGLE, Illustris, Illustris TNG and Horizon-AGN simulations. We now test whether our BCM implementation is able to recover such correlation.

In Fig.~\ref{fig:feedback} we show all possible power spectrum suppressions (within our prior parameter space) at $k=1 \ihMpc$ predicted by our model at a fixed baryon fraction. For comparison, stars show the mean BCM values found in the previous section for various hydrodynamical simulations. Also for comparison, dashed lines indicate an estimate for the largest possible suppression expected for a given baryon fraction, i.e. assuming haloes expel all their gas to infinity. Decomposing the power spectrum in dark matter and baryonic contribution, it is easy to show this is given by:

\begin{equation}
{\rm max} \left[ S(k) \right] = \left( 1-\frac{\Omega_b}{\Omega_m} \right)^2.
\label{eq:max_suppression}
\end{equation}

Firstly, we see that the BCM predicts a clear relation between $S(k)$ and the baryon content of clusters, including the relation reported in \cite{vandaalen2019} down to a baryon fraction $f_b\approx0.3$. For smaller baryon fractions, the predictions disagree. However, we note that in that regime \cite{vandaalen2019} relies on an extrapolation and indeed for $f_b\lesssim0.2$ predicts a larger suppression than the maximum expected.

We note that our relation is significantly looser than that of \cite{vandaalen2019} -- it is interesting to speculate the reasons behind this. On one hand, this could imply that there are fundamental relationships between the free parameters of the BCM, or that the functional forms provide more freedom than required. This could imply that a more deterministic model could be found in the future. On the other hand, many numerical simulations are calibrated to reproduce certain observables which might artificially limit the range of possible suppressions.

Very interestingly, the baryon fractions inferred by fitting the simulations (coloured stars) perfectly agree (< 1~\%) with the fitting function provided by \cite{vandaalen2019}, in all cases except for Illustris. This means that by only providing the clustering our model is able to correctly predict the amount of DM and baryons in simulated clusters. In the case of Illustris, on the other hand, our model indicates $f_b\approx10~\%$ the cosmic value, whereas the measurement from the hydrodynamical simulation is $f_b\approx35~\%$. Extreme feedback models, such as the Illustris one, appears to be strong enough to perturb the gas outside the halo boundaries. In order to reproduce the clustering of these simulations within the assumption of the model, i.e. no particle is displaced outside haloes, more gas needs to be expelled from the halo, resulting in an underestimation of the halo baryon fraction.

To confirm this hypothesis, we have fit another simulation of the Cosmo-OWLS suite, run with the same sub-grid implementation but with higher minimum heating temperature for AGN feedback, $T=10^{8.7}{\rm K}$.
We have found that in this case the baryon fraction is underestimated by a factor of $2$. Despite the fact that such strong feedback models are not preferred by observations, it will be interesting to explore further these aspects in the future, to refine the parametrisation and recipes of the BCM.

\subsection{Redshift evolution of baryonic parameters}
\label{sec:zevolution}

\begin{figure*}
  \includegraphics[width=0.4\linewidth]{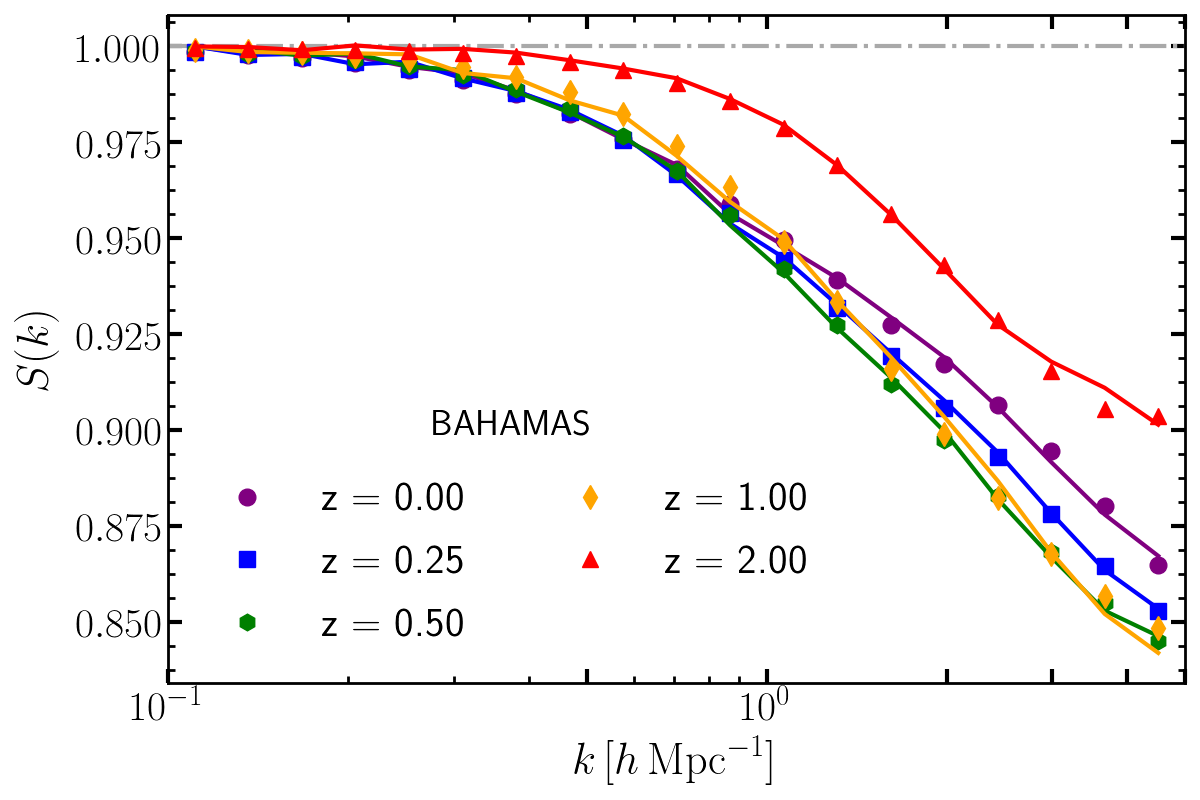}
    \includegraphics[width=0.4\linewidth]{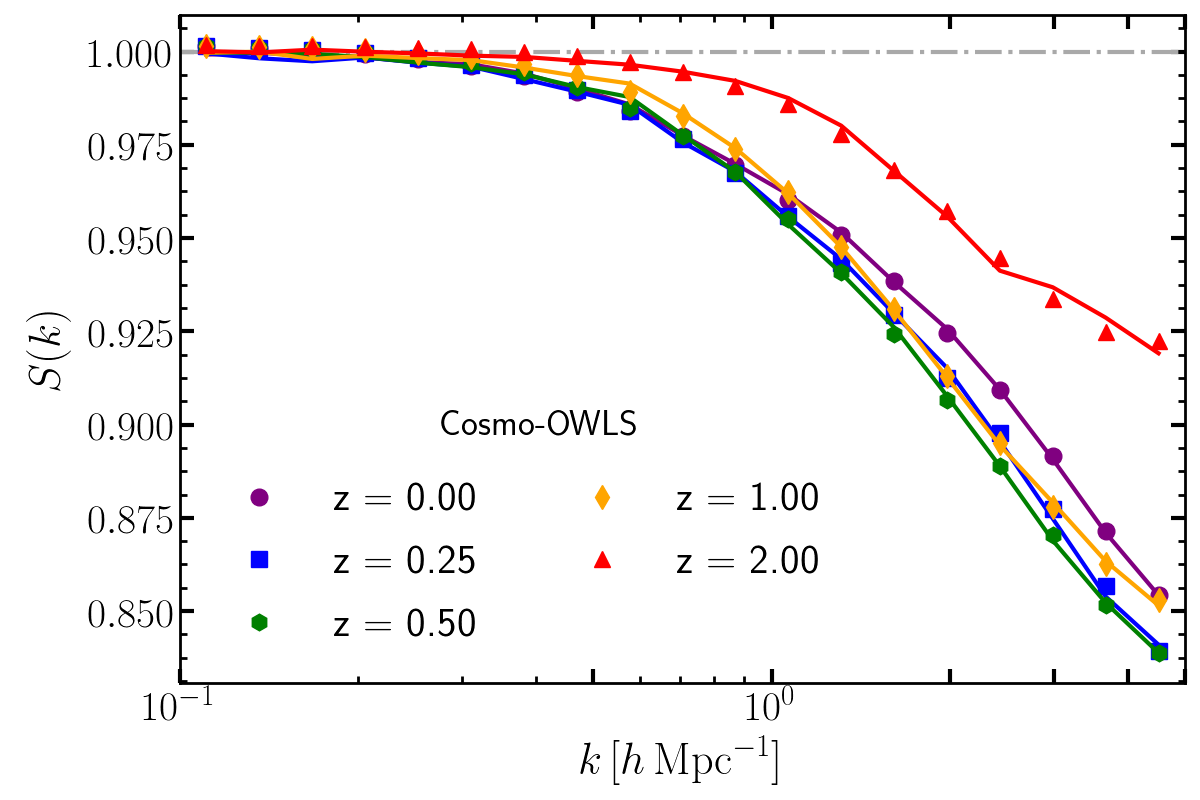} \\
    \includegraphics[width=0.4\linewidth]{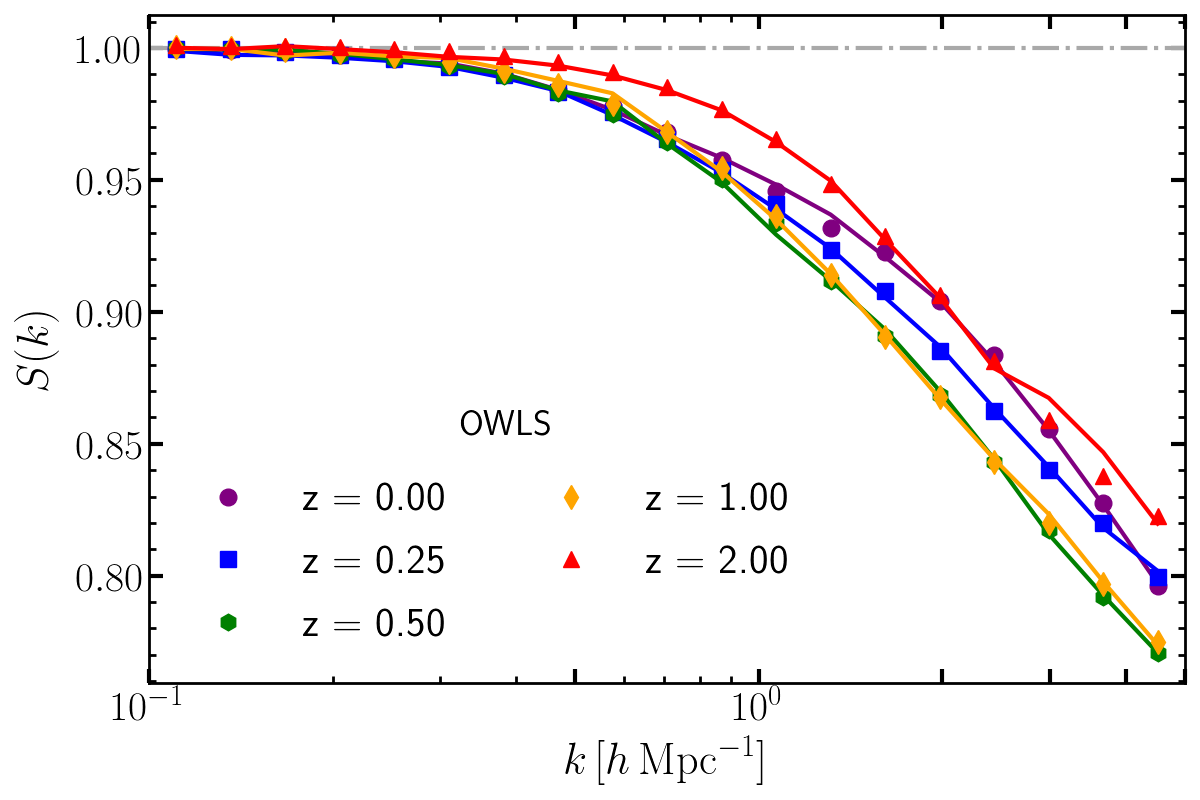}
  \includegraphics[width=0.4\linewidth]{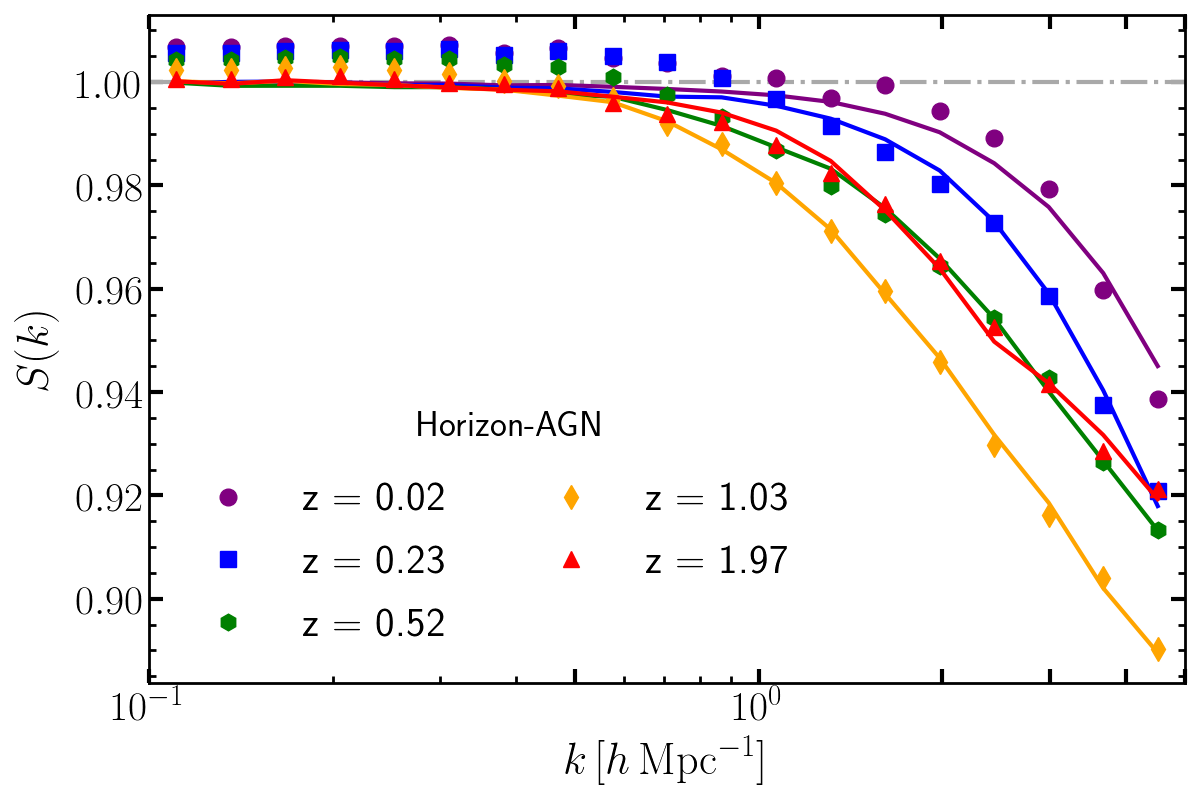}

  \caption{The impact of baryons on the power spectrum, $S(k) \equiv P/P_{GrO}$ as measured in the BAHAMAS (top left), Cosmo-OWLS (top right), OWLS (bottom left), and Horizon-AGN (bottom right) simulations at redshifts  0, 0.25, 0.5, 1, 2, as indicated by the legend. We do not show all the redshifts available for display reasons. Solid lines represent the best-fitting model.}
  \label{fig:bahamas_z_model}
\end{figure*}

Up to this point, we have only considered the baryonic effects at redshift zero. None of the BCM free parameters has a clear theoretical redshift dependence, except for $M_1$, for which we give a parameterisation based on halo abundance matching in Appendix \ref{app:profiles}. A naive approach would be to consider the other parameters constant, thus assuming that the evolution of the baryonic effects is only given by the evolution of the halo mass function. However, it has already been proven that this is not the case. The BCM fitting function parameters provided by \cite{S&T2015} show in fact a clear redshift dependence, when applied to Horizon-AGN at different snapshots \citep{Chisari2018}. In this section, we extend \cite{Chisari2018} analysis by fitting the power spectrum suppression for BAHAMAS, Cosmo-OWLS, OWLS and Horizon-AGN at multiple redshifts between $0 \le z \le 2$. We perform the fit with the same setup used in the previous section for $z=0$.

We display the measured $S(k)$ along with the best-fitting BCM predictions in Fig.~\ref{fig:bahamas_z_model}. Firstly, we can see a clear evolution of $S(k)$, with an amplitude that is typically smaller at high $z$. This is comparable with the analysis of \cite{Chisari2018}.
Remarkably, our model provides an excellent fit for the data over all the scales and redshifts considered, achieveng a percent accuracy even in the most extreme cases.

\begin{figure*}
  \includegraphics[width=0.65\linewidth]{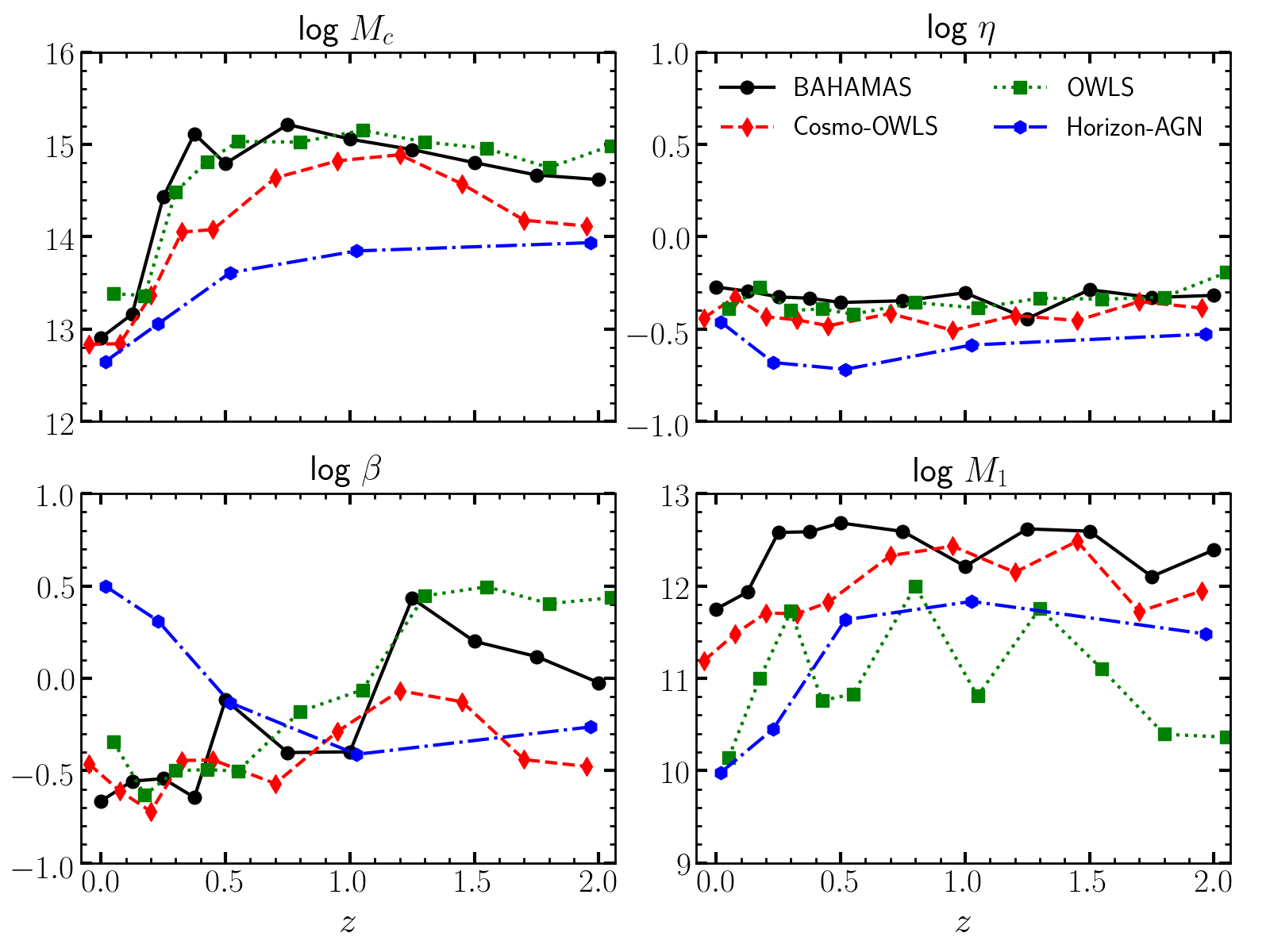}
  \caption{Marginalised values of the best-fitting BCM parameters for the BAHAMAS, Cosmo-OWLS, OWLS, and Horizon-AGN simulations at different redshifts $0 \ge z \ge 2$, as indicated by the legend. Note that the redshifts of the measurements are slightly shifted for display purposes.}
  \label{fig:bahamas_z_parameters}
\end{figure*}

In Fig.~\ref{fig:bahamas_z_parameters} we show the expectation values for $M_c$, $\eta$, $\beta$, and $M_1$ as a function of redshift. We find that BAHAMAS, Cosmo-OWLS and OWLS do not show a significant evolution of the AGN feedback range, having the $\eta$ parameter roughly constant in time. On the contrary, Horizon-AGN shows a monotonic increase of $\eta$, in agreement with the finding of \cite{Chisari2018}. The power spectrum suppression $S(k)$ of the hydrodynamical simulations in study roughly peaks around $z\approx1$. Therefore, for the correlation shown in Fig.~\ref{fig:feedback} we can expect the peak of the quantity of gas expelled from haloes around this redshift.

Indeed, we find that the mean values of $M_c$ in all the simulations increase up to $z=1$ and a slowly decrease afterwards, except for Horizon-AGN in which $M_c$ monotonically increases. The characteristic host halo mass $M_1$ shows a similar trend, increasing at low redshifts and staying somewhat constant after $z\approx0.5$. Finally, it appears that for OWLS, Cosmo-OWLS and BAHAMAS steeper transitions in mass from gas-rich to gas-poor haloes are preferred at higher redshifts. At odds with this trend, Horizon-AGN mean values of $\beta$ monotonically decrease from $z=0$ to $z=1$.

In conclusion, it is clear that it is required a large BCM parameter space in order to describe different state-of-the-art hydrodynamical simulations. The redshift evolution of the parameters shows some similarities but it is not always consistent among the simulations. All this emphasises the importance of having flexible and general recipes in the BCM, at the risk of biasing parameter estimates.

\section{Information analysis: baryon-cosmology degeneracies}
\label{sec:fisher}

In the previous sections we have shown that our framework can simultaneously model cosmology and astrophysics in the mass power spectrum. Now, we explore the degeneracies between them, and investigate how much cosmological information is lost after marginalising over the free parameters of the BCM.

\subsection{Fisher Matrix}

We employ a Fisher formalism to quantity the amount of information encoded in the mass power spectrum. Notice that we refrain from modelling the shear power spectrum (which would correspond to a convolution of the mass power spectrum with the relevant lensing kernel) to keep our study as general and independent of details of a particular experiment (e.g. the redshift distribution of background galaxies) as possible.

Using the power spectrum $P(k)$ as our observable, and assuming a multivariate Gaussian distribution, the Fisher matrix is defined as:

\begin{equation}
\mathcal{F}_{ij} \equiv \frac{\partial P}{\partial \vartheta_i} \mathcal{C}^{-1} \frac{\partial P^\dagger}{\partial \vartheta_j}
+ \frac{1}{2} tr \left[  \mathcal{C}^{-1} \frac{\partial  \mathcal{C}}{\partial \vartheta_i} \mathcal{C}^{-1}
\frac{\partial \mathcal{C}}{\partial \vartheta_j} \right]
\label{eq:FIM}
\end{equation}

\noindent where $\mathcal{C}$ is the observable covariance matrix. We neglect the second term of Eq.\ref{eq:FIM} to ensure the conservation of the information \citep{Carron2013}, noting however that $\mathcal{C}$ depends very weakly on cosmology and that term would be negligible \citep{Kodwani2019}.

\subsubsection{Model parameters and priors}

Our fiducial model will consist of a 8-parameter cosmology: five parameters describing a minimal model ($\Omega_m$, $\Omega_b$, $h$, $n_s$, $A_s$), one parameter describing the total neutrino mass ($\sum m_{\nu}$), and two parameters describing the dark energy equation of state, $w_0$ and $w_a$ in the Chevallier-Polarski-Linder parametrisation, \citep{ChevallierPolarski2001,Linder2003}. We assume fiducial values for these parameters consistent with the current constraints from CMB+BAO+Lensing \citep[][hereafter Planck18]{Planck2018}. Specifically: $\Omega_{cdm}=0.261$, $\Omega_{b}=0.04897$, $\Omega_{\Lambda}=0.69889$, $H_0=67.66 \,  {\rm km  \,s}^{-1} \hMpc$, $n_s=0.966$, $A_s=2.105 \times 10^{-9}$, $w_0=-1$, $w_a=0$, $\sum m_{\nu}=0.06$~eV.

We will also consider 4 additional baryonic parameters specifying the BCM. In particular, the best-fitting values of BAHAMAS found in \S\ref{sec:hydro_bcm}: $M_c=3.3\times10^{13} ~ \Msun$, $\eta=0.54$, $\beta=0.12$, $M_1=8.63\times10^{11} \Msun$. This specific choice is justified by noticing that the BAHAMAS simulation have been specifically calibrated to match the observed baryon fraction in haloes, a quantity that is well correlated with baryonic clustering effects.
Therefore, we expect its predictions to be more reliable for this analysis. Moreover, the cosmological framework of the simulation, which is given by Planck 2015 best-fitting and includes massive neutrinos, is very similar to our fiducial one. We set the redshift of our analysis at $z=0.25$, around which the lensing window of most of the current and forthcoming lensing surveys is peaked.

\subsubsection{Covariance matrix}

Very often, when computing the covariance matrix of the observable, $\mathcal{C}$, it is implicitly assumed a perfect theoretical model over the whole range of scales. This, however, is not correct in general. Specifically, for the case of nonlinear power spectrum, there are model uncertainties (arising from, for instance, how baryonic effects are described, the solution of the Vlassov-Poisson equations by $N$-body simulations, or emulation uncertainties) that should be taken into account. Therefore, we split our covariance matrix in two terms: $\mathcal{C}=\mathcal{C}_{\rm D}+\mathcal{C}_{\rm T}$, where $\mathcal{C}_{\rm D}$ describes the data covariance and $\mathcal{C}_{\rm T}$ the theory one. We employ a Gaussian data covariance which reads

\begin{equation}
\mathcal{C}_{{\rm D},ij} = \delta_{ij}\,\frac{2}{N_k} \left[ P(k_i) + \frac{1}{\bar{n}} \right]^2,
\label{eq:data_covariance}
\end{equation}

\noindent where $N_k$ is the number of independent modes in each bin, approximated as $N_k=V_{\rm box}k^2\delta k / (2\pi)$, and $1/\bar{n}$ is the shot noise term. The reference power spectrum $P(k)$ is computed with {\tt halofit} \citep{Takahashi2012} within the fiducial cosmology Planck18, and we consider for the shot noise term a total volume of $1 \, h ^{-3}{\rm Gpc}^3$  and a number density $\bar{n}=5\cdot 10^{-2}~h^3$~Mpc$^{-3}$.

For $\mathcal{C}_{\rm T}$ we employ the same procedure used in \S\ref{subsec:bcm_pars}, Eq.~\ref{eq:suppression_covariance}.
We consider here as sources of model error the BCM and the cosmology rescaling. For the first, we assume $\mathcal{E}$ to be a constant with amplitude $1\%$ of the reference power spectrum, and a correlation length $\ell=1\,\ihMpc$ which is motivated by our findings in Fig.~\ref{fig:hydro_models}.

For the term originating from the cosmology rescaling, $\mathcal{E}$ is a constant $P(k)/100$ that raises smoothly up to 2~\% at $k\sim1\,\ihMpc$, $\mathcal{E}=[1.5+0.5\,\erf(k-1)]P(k)$, and same correlation length $\ell=1\,\ihMpc$, since typical deviations from the target simulation are similar on all scales (c.f. Fig.~\ref{fig:pk_scaling} and \citealt{Contreras2020}). The theory covariance is simply given by the sum of the two contributions described above.

\subsubsection{Numerical derivatives}

\begin{figure}
  \includegraphics[width=\linewidth]{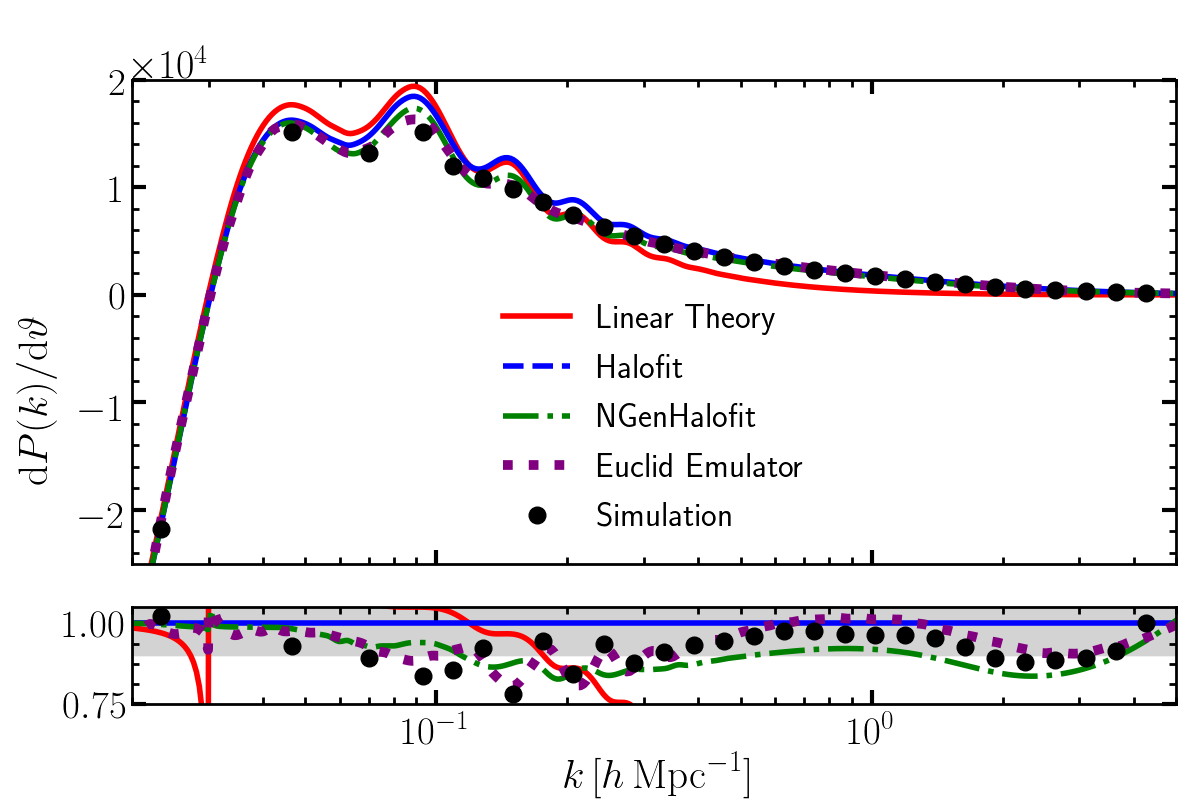}
  \caption[test caption]{ {\it Upper panel:} derivatives of the matter power spectrum at $z=0.25$ with respect to $\Omega_{\rm cdm}$, evaluated for a Planck18 (with $\sum m_{\nu}=0 \, {\rm eV}$) cosmology and a BAHAMAS-like baryonic model. The black symbols indicate the results obtained using our cosmology scaling technique, which we compare against linear theory (red), {\tt halofit} (blue), {\tt NGenHalofit}(green) and {\tt EuclidEmulator} (purple). {\it Lower panel:} ratio over {\tt halofit} of the derivatives shown in the upper panel.}
  \label{fig:derivatives_comparison}
\end{figure}

\begin{figure*}
  \includegraphics[width=0.9\linewidth]{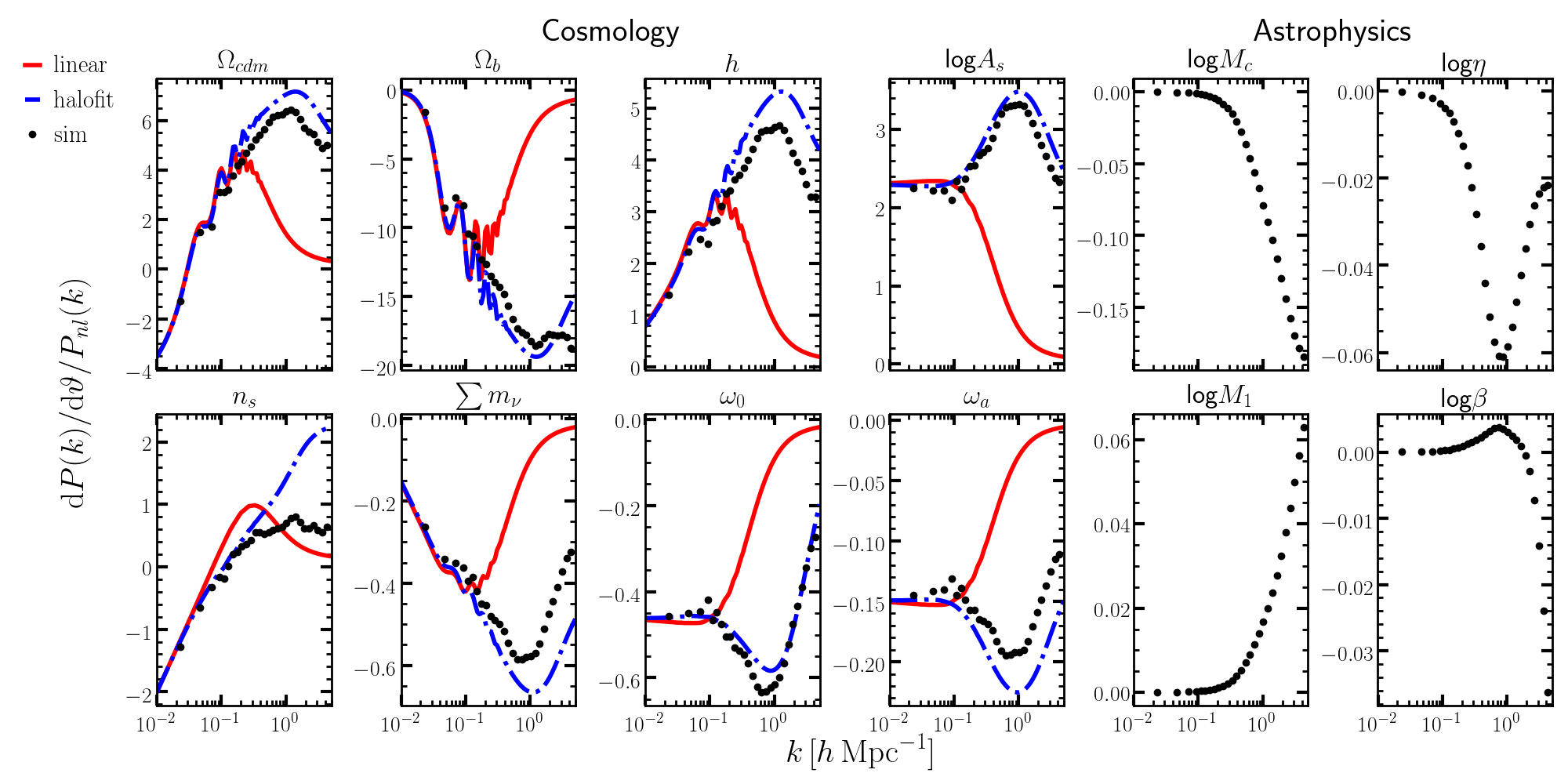}
  \caption{Derivatives of the matter power spectrum around the cosmology preferred by \citep{Planck2018} at $z=0.25$. Symbols display the results computed with rescaled $N$-body simulations, whereas blue and red solid lines do so for {\tt halofit} and linear perturbation theory, respectively.}
  \label{fig:derivatives}
\end{figure*}

The next ingredient for computing the Fisher matrix elements is the estimation of the partial derivatives $\partial P/\partial \vartheta_i$. We compute these using second-order-accurate central finite differences:

\begin{equation}
\frac{\partial P_{\vartheta}(k)}{\partial \vartheta}\approx\frac{P_{\vartheta + \epsilon}(k) - P_{\vartheta - \epsilon}(k)}{2\epsilon}.
\label{finite_difference}
\end{equation}

\noindent We have checked that using the fourth-order approximation the results are practically identical. The parameter intervals, listed in Tab \ref{tab:intervals}, are chosen to produce a 1~\% effect in the matter power spectrum in the range $[0.01,5]\,\ihMpc$.
We have carefully checked that these intervals are sufficiently small so that the power spectrum response is still linear but large enough so numerical noise is reduced.

Operationally, we rescale the cosmology of our simulation to the required parameter set, apply the BCM, and then measure the power spectra. In this analysis we use a set of paired simulations run with $768^3$ particles and a box size of $256 \hMpc$ described in \S\ref{sec:sim}. To increase the precision of the scaling, we furthermore apply the interpolation between snapshots discussed in section \S\ref{sec:cosmoscaling}. Therefore, for each point in the parameter space we scale the cosmology of four snapshots (two paired), and apply the BCM four times. In this case, we measure the power spectrum in bins which are multiples of the fundamental mode.

To achieve the extremely high precision required by the Fisher matrix calculation (e.g. numerical stability of the matrix inversion) we make a regression of our data in logarithmic bins for $k>0.1 \ihMpc$, using the Gaussian processes framework {\tt Gpy} \citep{gpy2014}, and furthermore applying a gaussian smoothing to remove any residual small-scale noise.

To test the accuracy of our results we compare our derivatives against those predicted by linear theory given by the Boltzmann solver {\tt CLASS} \citep{Lesgourgues2011}, {\tt halofit} \citep{Takahashi2012}, {\tt EuclidEmulator} \citep{EuclidEmulator} and {\tt NGenHalofit} \citep{Smith&Angulo2019}. The {\tt EuclidEmulator} is built using a suite of 100 simulations run with different cosmologies, with a nominal absolute accuracy of 1\%, whereas {\tt NGenHalofit} is a 3\%-accurate extension of {\tt halofit} obtained by calibrating against the D\"{a}emmerung suite of simulations. Since none of these two codes support massive neutrino cosmologies, we perform the comparison assuming $\sum m_{\nu}=0 \, {\rm eV}$ and furthermore neglect baryonic effects.

In Fig.~\ref{fig:derivatives_comparison} we show our results for $\vartheta=\Omega_{cdm}$. On the largest  scales considered, {\tt halofit}, {\tt NGenHalofit}, {\tt EuclidEmulator} and our cosmology scaling technique all perfectly agree. On intermediate scales ($k>0.1\,\ihMpc$) the methods start to disagree at the $20\%$ level. Specifically, our method, {\tt EuclidEmulator} and {\tt NGenHalofit} are in very good agreement but are systematically different from linear theory and {\tt halofit}.

It is also interesting to note that BAO oscillations in {\tt NGenHalofit} are damped more efficiently with respect to {\tt halofit} and linear theory, but not as much as in {\tt EuclidEmulator}. On small scales, {\tt EuclidEmulator} predictions depart from those of {\tt NGenHalofit}, providing again similar results around $k\approx4 \,\hMpc$. The cosmology scaling algorithm predicts power spectra which match the ones from {\tt EuclidEmulator} within 1\%, and accordingly the derivatives appear to be at the same accuracy level, supporting the validity of our approach. Although not shown here, we have checked that we obtain similar conclusions when considering other cosmological parameters in our set. It is also worth to highlight that we have obtained our results with only two relatively small simulations, $L=256\ihMpc$.

\begin{table}
  \centering
  \begin{tabular}{l|c}
     \hline
     parameter ($\vartheta$) & interval ($\epsilon$) \\
     \hline
     $\Omega_{cdm}$  & $2.6 \times 10^{-3}$ \\
     $\Omega_{b}$    & $7.0 \times 10^{-4}$ \\
     $H_0 \, \,[ { \rm Km\,s^{-1} Mpc^{-1}}]$  & $1.6 \times 10^{-3}$ \\ 
     $n_s$           & $3.5 \times 10^{-3}$ \\
     $\log \, A_s$   & $2.7 \times 10^{-3}$ \\
     $w_0$           & $5.1 \times 10^{-2}$ \\
     $w_a$           & $5.2 \times 10^{-2}$ \\
     $\sum m_{\nu} \, [{\rm eV}]$  & $2.0 \times 10^{-2}$ \\
     $M_c \, \, [\Msun]$         & $1.2 \times 10^{-1}$ \\ 
     $M_1 \, \, [\Msun]$           & $3.5 \times 10^{-1}$\\  
     $\eta$          & $1.3 \times 10^{-1}$ \\
     $\beta$         & $1.4 \times 10^{-1}$ \\
     \hline
     \end{tabular}
    \caption{Parameter intervals used to compute numerically the derivatives of the power spectrum. The values were chosen to cause 1\% change in the nonlinear matter power spectrum over the range $k\in [0.01-5]\,\ihMpc$.}
  \label{tab:intervals}
\end{table}

Having tested our implementation against other nonlinear models, we now consider our entire parameter space including massive neutrino and dynamical dark energy. In Fig.~\ref{fig:derivatives} we show the measured partial derivatives with respect to each of our 12 parameters. Linear theory and {\tt halofit} predictions are overplotted for reference in the case cosmological parameters.

Cosmological derivatives provided by the three methods agree on large scales but differ on smaller scales ($k>0.1 \ihMpc$). The good agreement between our approach with {\tt NGenHalofit} and {\tt EuclidEmulator} shown in Fig.~\ref{fig:derivatives_comparison} suggests the differences arise from inaccuracies in {\tt halofit} rather than in the cosmology rescaling.

The dependence of the power spectrum with BCM parameters is consistent with that shown in Fig.~\ref{fig:bcm_pk_dependence}, being the range of the AGN feedback parametrised by $\eta$ the one impacting the power spectrum on larger scales, while the galaxy formation parametrised by $M_1$ causes an enhancement of power on small scales.

\subsection{Information in the mass power spectrum}

In the Gaussian approximation, the total amount of information that we can extract up to a given scale is proportional to the number of independent modes contained in that scale. Since the number of independent modes $N_{k}$ goes as $N_k \propto k^3$, it is evident that even a modest increase of the smallest scale modelled can unlock a big amount of information. Small scales, on the other hand, are more affected by baryonic physics. In this subsection we will explore this interplay.

\begin{figure*}
  \includegraphics[width=0.8\linewidth]{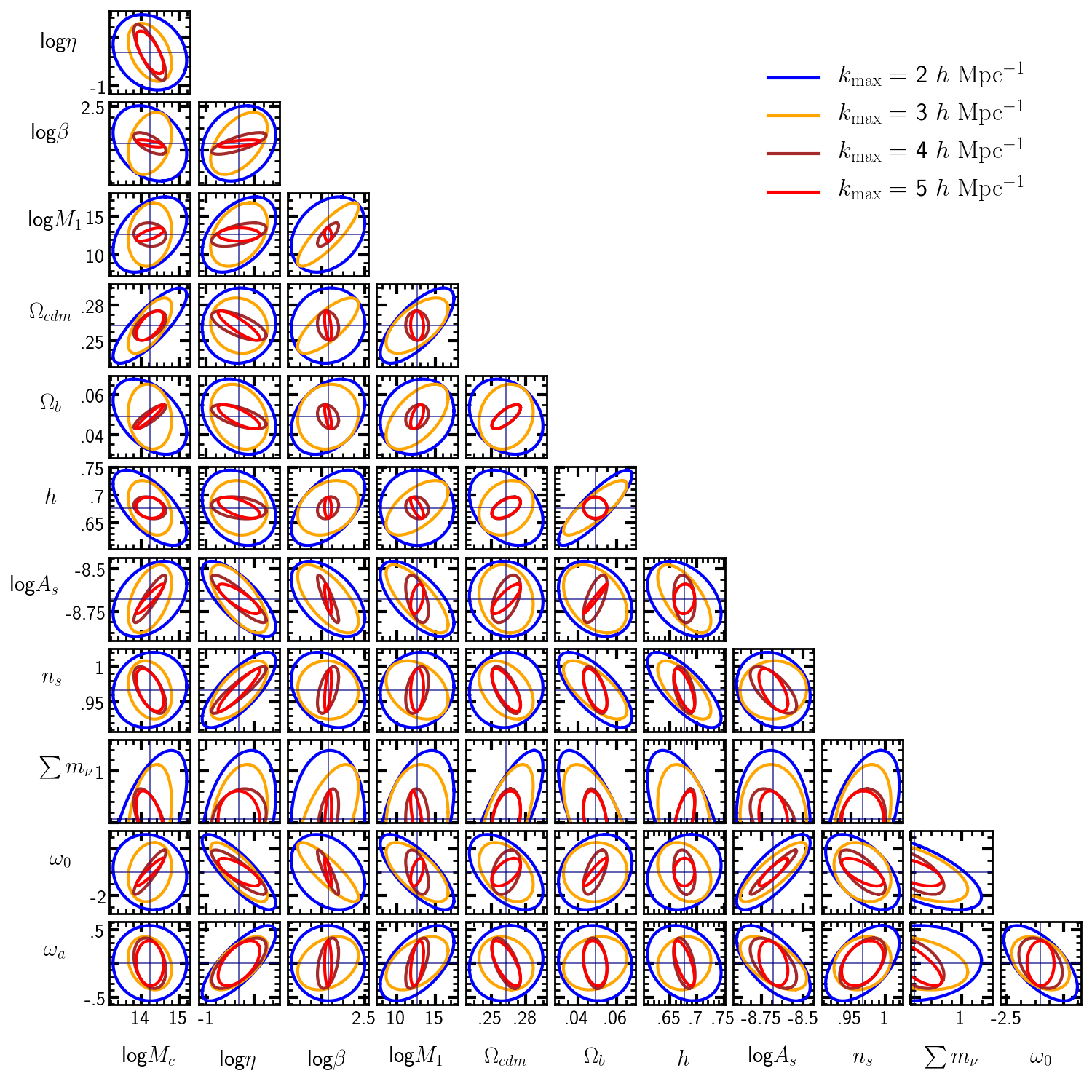}
  \caption{$1\sigma$ ellipses computed considering as maximum wavenumber of the analysis $k_{\rm max}$=2,3,4 and 5 $\ihMpc$ (blue, orange, brown and red solid lines, respectively).  These Fisher forecasts have been computed using a
$256 \, \hMpc$ simulation, scaled to Planck18 cosmology and employing a BAHAMAS-like baryonic feedback at redshift $z=0.25$. Notice that we are not considering the theoretical contribution to the covariance matrix to show the dependence of the parameter degeneracies with the minimum scale.
            }
  \label{fig:fisher_kmax}
\end{figure*}

\noindent In Fig.~\ref{fig:fisher_kmax} we show the $1\sigma$ credibility regions of baryonic and cosmological parameters, employing as minimum scales wavenumbers from $2 \,\ihMpc$ to $5\,\ihMpc$. The baryonic parameters show many degeneracies, both between each other and the cosmological parameters. In particular, for large $M_c$ the model prefers low $\eta$ and $h$, and large values of $w_0$, $A_s$, $\Omega_{cdm}$ . The $\beta$ parameter is degenerate with $M_1$, and for large values of $\eta$ are preferred large $M_{1}$.
Moreover, $\eta$ is degenerate with the dark energy equation-of-state parameters, preferring large values for low $w_0$ and high $w_a$.

As we consider smaller scales, constraints improve and some of the degeneracies flip direction or are completely broken. For example, large values of $h$ seem to prefer high $M_{1}$ considering only scales up to $k=3\, \ihMpc$, but low $M_{1}$ extending the analysis to smaller scales. The degeneracy between $h$ and $\Omega_b$ is broken including scales $k \ge4\,\ihMpc$. At large scales, the sum of neutrino masses $\sum m_{\nu}$ shows an anticorrelation with $M_c$, a result similar to what  found by \cite{Parimbelli2019}, using \texttt{halofit} combined with BCM fitting function up to $k\le0.5 \, \ihMpc$ in weak lensing forecasts. We note that, however, including the small scales the situation changes, and for large neutrino masses are preferred high values of $M_c$.

\begin{figure*}
  \includegraphics[width=\linewidth]{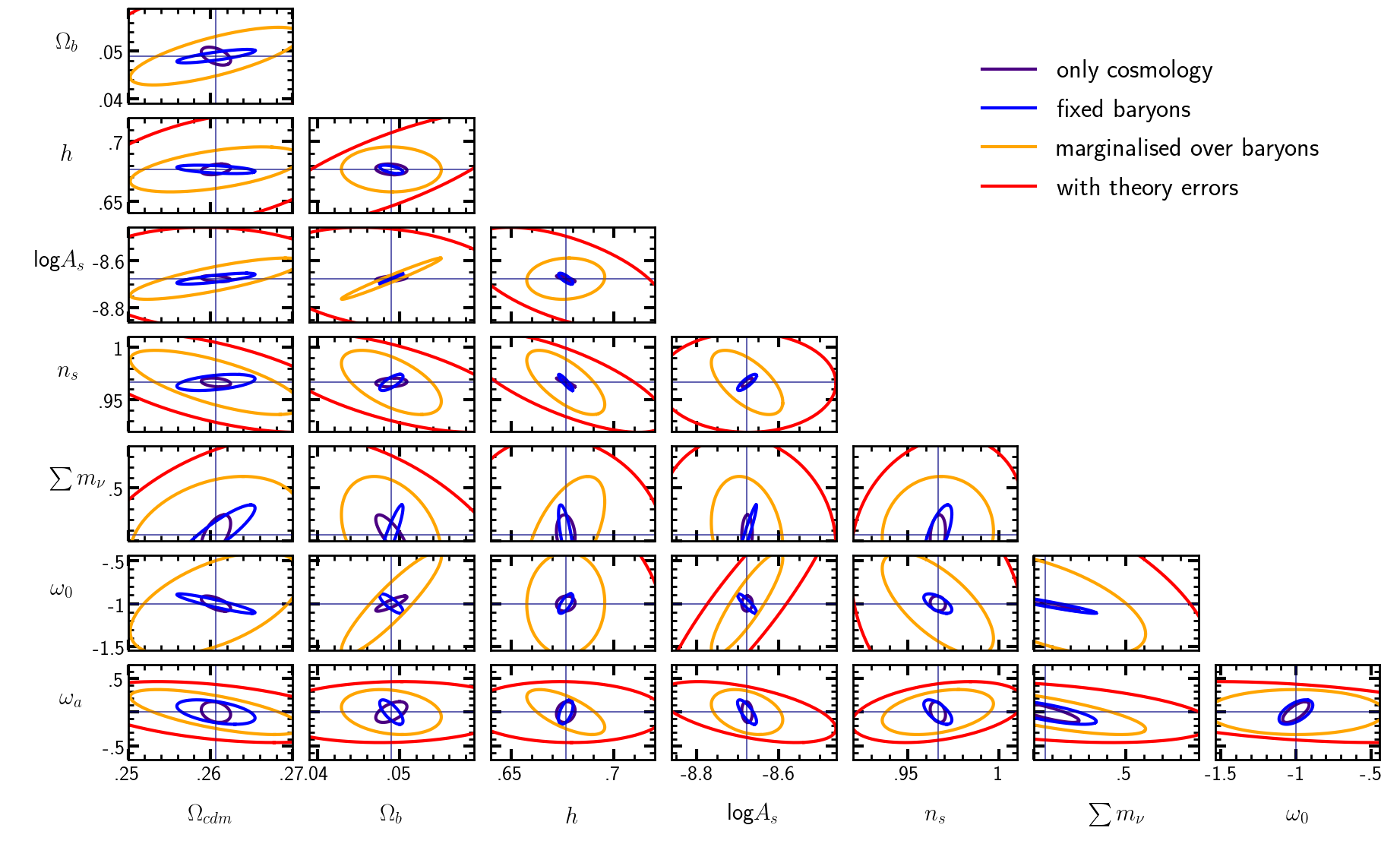}
  \caption{$1\sigma$ Fisher contours of the matter power spectrum measured at scales $k\le5 \, \ihMpc$, considering only cosmology (purple line), an exact baryon modelling (blue), marginalising over the baryonic parameters (orange), and including in the marginalisation the theory errors (red).}
  \label{fig:fisher_cfm}
\end{figure*}

\noindent In principle, it is possible to constrain the BCM parameters through observations, e.g. the halo baryonic fraction and stellar-to-halo mass relation through X-ray, thermal Sunyaev-Zeldovich effect or weak-lensing data. In practice, however, the actual constraints on the BCM parameters are loose because of both observational and modelling uncertainties \citep[e.g. hydrostatic mass bias, see][]{S&T2018}.

\begin{figure}
  \includegraphics[width=\linewidth]{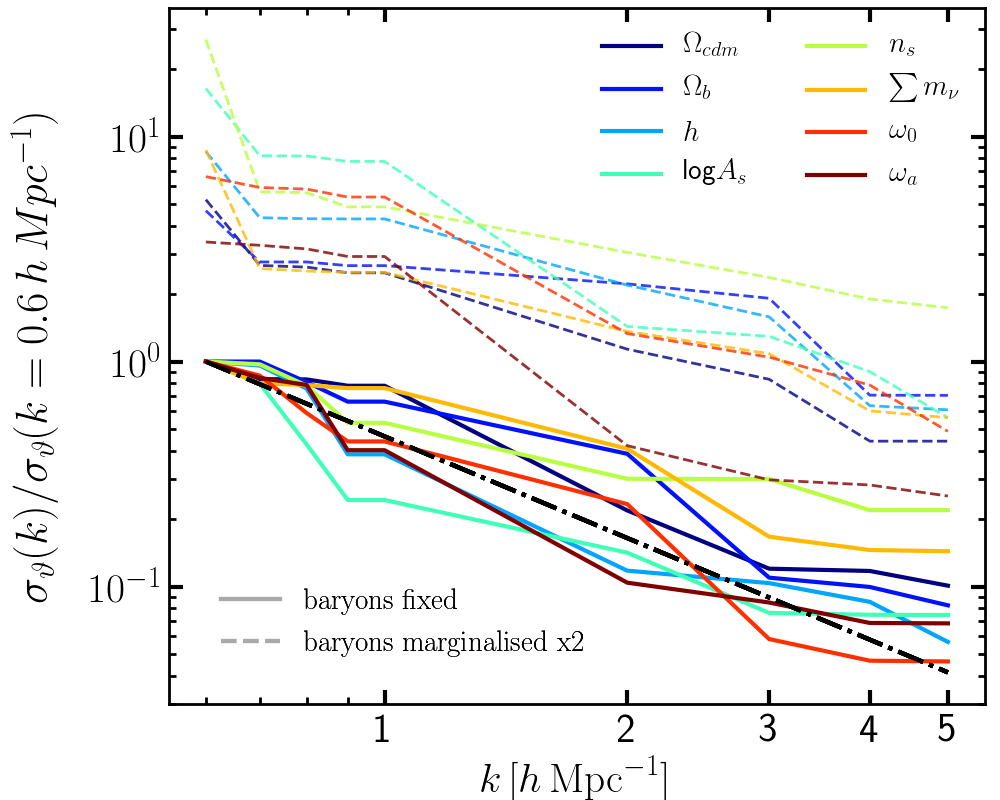}
  \caption{Expected accuracy in cosmological parameters constraints as a function of the maximum wavenumber $k$ used in the analysis. Different colours display the results for different parameters, as indicated by the legend. Dashed lines show the ratio between results obtained marginalising and fixing baryonic parameters respectively. Note the former are multiplied by a factor of two for display purposes. The dashed-dotted line represents the ideal scaling expected for Gaussian fields, $\propto k^{-3/2}$.}
  \label{fig:fisher_scaling}
\end{figure}

\noindent A viable alternative can be the marginalisation over the baryonic parameters
to avoid biased results in the estimation of the cosmological parameters, at the price of loosing constraining power.
In Fig.~\ref{fig:fisher_cfm} we show the $1\sigma$ credibility regions obtained by ignoring, fixing and marginalising over the baryonic physics. Interestingly, the degeneracies of the cosmological parameters slightly change if we consider or not the baryonic effects, even if we do not marginalise over them.
Notice that by construction the ellipses are centred about the true values of
the parameters, therefore the plot is not meant to show the possible biases of the parameters estimation. It is also important to note that, despite the use of a state-of-the-art modelling within high resolution $N$-body simulations, the theoretical errors of the model are still the main uncertainties on the constraints, and must be incorporated in each pipeline to avoid bias in parameter estimations. On the other hand, we find that the marginalisation over the baryonic parameters have a quite different impact on the constraining power for different parameters.

In Fig.~\ref{fig:fisher_scaling} we display the expected marginalised $1\sigma$ constraints in the parameters employing the information up to varying wavelengths. We display a case where we assume perfect knowledge of the astrophysical processes (solid lines) and where we marginalise over the baryonic parameters (dashed lines). It is evident that the impact of the marginalisation is scale-dependent, and generally larger on large scales. The constraints obtained at $k=5\,\ihMpc$ for $\sum m_{\nu}$, $\Omega_b$ and $w_a$ are factors of $\sim2$ larger after the marginalisation. On the contrary, $h$, $A_s$ and $w_0$ have a factor of $\sim4-5$ weaker constraints, with the other parameters falling in between.

Recently, \cite{Schneider2019} have performed a similar study, fitting the shear power spectrum using a model which combines the predictions of {\tt halofit} with an emulator built upon a baryon correction model. They find that constraints in $n_s$ are more than a factor of $2$ weaker after the marginalisation, whereas $h$ and $\Omega_m$ less than 50\%. These results are in broad agreement with our findings, even if a more direct comparison is not possible because of the different assumptions in the BCM setup, the characteristics of the target survey and the observable used in the analysis.

\section{Discussion and Conclusions}
\label{sec:discussion}

Cosmological observations are entering an age where uncertainties in data models are significantly limiting the inferred parameter constraints. In particular, for weak gravitational lensing, not only the nonlinear evolution of density fluctuations but also details of galaxy formation theory and the evolution of cosmic gas become important for correctly interpreting future measurements.

Jointly modelling cosmology and astrophysics via hydrodynamical simulations requires a huge computational effort because of the large dynamical range and number of (still uncertain) astrophysical processes involved. In addition, hydrodynamical simulations make specific choices about the physics included, hydrodynamical solver, and the free parameters of each sub-grid recipe. To consistently and systematically explore different models of both cosmology and astrophysics appears simply unfeasible with the current computational power without making multiple assumptions.

In this context, baryonic correction and cosmology rescaling methods appear to provide a fast and flexible approach to capture the effects of astrophysics and cosmology on the full density field. The main idea is to model the impact of baryonic physics with a minimal set of recipes (motivated by observations and numerical simulations), which are applied in post-processing to a gravity-only simulation with varying cosmologies, as provided by cosmology-rescaling methods. In this way, observations such as weak-lensing and Sunyaev-Zeldovich could simultaneously constrain cosmological and astrophysical parameters.

Key points of this method are the dispensable use of a full set of $N$-body simulations, the relatively easy incorporation of extensions to $\Lambda$CDM, the large parameter space covered, and the possibility of carrying out larger and more accurate simulations to achieve a higher precision in the matter clustering measurements. We recall moreover that for each of the baryon-cosmology set we obtain a full 3D prediction for the galaxy, gas, and star distributions, opening up interesting possibilities of predicting the cross-correlations of different observables. \\
In this paper we have discussed one possible implementation of such baryonic correction models applied on one paired $N$-body simulation rescaled to different cosmologies. Below we summarise the main findings:

\begin{itemize}
 \item Our specific baryon correction model (BCM) is able to describe the mass matter power spectrum up to $k=5\ihMpc$ at $z=0$, achieving an accuracy of $<1\%$ for all 7 state-of-the art hydrodynamical simulations here considered (Fig.~\ref{fig:hydro_models}).

 \item By only fitting the mass clustering, we are able to recover the correct halo baryon fraction in most of the cases, except for extreme feedback models (e.g. Illustris, Fig.~\ref{fig:feedback}).

 \item Different hydrodynamical simulations prefer different values and redshift evolution for the free parameters of the BCM (Fig.~\ref{fig:hydro_contours}). Despite this, there is a relatively tight correlation between the baryon fraction in clusters and the baryon-induced power spectrum suppression (Fig.~\ref{fig:bahamas_z_parameters}).

  \item Applying our BCM to cosmology-rescaled simulations adds only $<1\%$ uncertainty to the whole approach (Fig.~\ref{fig:pk_scaling}).

 \item Using a Fisher matrix formalism we explore the impact of baryons on the information available in the mass power spectrum up to $k\sim5\ihMpc$ (Fig.~\ref{fig:derivatives}). We find baryons change the sensitivity of $P(k)$ to cosmology, altering the degeneracy among parameters.

 \item After a marginalisation over the free parameters of the BCM, there is a moderate degradation of constraining power (see Fig.~\ref{fig:fisher_scaling}). Specifically, constraints decrease by factors of 2-4 depending on the parameter considered. Naturally, these values will depend on the specific setup of a given survey.

  \item Errors and uncertainties in any data model exist and cannot be neglected. We estimate the magnitude of such uncertainty for our approach to be of about $\sim2\%$. We incorporate these errors in our Fisher analysis and find that it degrades cosmological information further than a marginalization of the BCM free parameters (red contours, Fig~\ref{fig:fisher_cfm}) for the setup considered here.

\end{itemize}

\noindent In conclusion, the combination of BCM and cosmology rescaling provides a powerful framework for simultaneously model cosmology and baryonic physics for large scale structure, and weak gravitational lensing in particular. There are some aspects that are important to highlight in the context of future data analyses.

First, it is crucial to understand and quantify all sources of uncertainties in data models. This can be achieved, for instance, by analysing mock lensing surveys constructed from state-of-the-art hydrodynamical simulations and realistic conditions. Naturally, the exact impact of theory errors depend on the specifics of the target observational setup and the model employed, but we emphasise the need of more flexible rather than more deterministic models for baryonic effects.

Second, it is important to guarantee that the BCM is both minimal and flexible enough so that derived constraints are as tight as possible, but also that they do not make strong assumptions regarding the underlying baryonic physics. This is crucial to ensure robust constraints on, for instance, neutrino masses. Thanks to the clear physical meaning of each component and the free parameters of the BCM, the assumptions, functional form and values can be directly compared with hydrodynamical simulations under extreme physics implementations.

Third, the accuracy of the BCM, the cosmology rescaling, and numerical simulations as a whole can be improved further. For instance, better concentration-mass relation models and the joint use of various simulations can improve the accuracy of cosmology-rescaling; systematic comparison among different $N$-body codes as a function of force and time-integration accuracy; and direct testing of the recipes within the BCM.

\noindent An interesting path is to extend the BCM to other gas properties. This would  pave the way for joint analysis of multiple observables. For instance, the extent of the expelled gas should not only affect lensing observables, but also the amount of gas detectable via Sunyaev-Zeldovich effect. Similarly, the amount of galaxies and their mass affects the lensing signal on small scales, but also affects the galaxy correlation function on small scales. The use of high-resolution simulations enabled by the cosmology-rescaling should also allow for more sophisticated and realistic modelling of the galaxy-halo connection, which should ultimately improve the performance of the BCM and reduce free parameters. We plan to explore all this in future works.

\label{sec:conclusion}

\section*{Acknowledgements}

The authors acknowledge the support of the E.R.C. grant 716151 (BACCO). C.H.-M. acknowledges support from the Spanish Ministry of Economy and Competitiveness (MINECO) through the projects AYA2015-66211-C2-2 and PGC2018-097585-B-C21.
SC acknowledges the support of the ``Juan de la Cierva Formaci\'on'' fellowship (FJCI-2017-33816).
We thank Elisa Chisari, Wojtek Hellwing, Volker Springel, Marcel van Daalen and Mark Vogelsberger as well as the Horizon-AGN and OWLS collaborations for kindly making various simulation data available to us.
We thank Bj\"{o}rn Malte-Schaefer and Marcel van Daalen for useful comments on the draft.
We acknowledge Louis Legrand for providing useful functions for the Fisher analysis.



\bibliographystyle{mnras}
\bibliography{bibliography} 



\appendix

\section{Analytical profiles of baryons and dark matter}
\label{app:profiles}
In this appendix we report the analytical formulae used in this work to compute the
density profiles of each component of the BCM. In the end, we
obtain a total ``baryon corrected''  mass profile that we have to invert in order to find
the lagrangian displacement required to make the halo particles match the new profile.
For every halo of the simulation, we fit a NFW profile, and we model our initial, gravity-only
 density profile as follow:

\begin{equation}
\rho_{\rm GrO}(x) =  \left\{ \begin{array}{cc}
        \rho_0 \cdot x^{-1} (1+x)^{-2} & \hspace{5mm} r \le r_{200} \\
                  0 & \hspace{5mm}  r > r_{200} ,\\
                \end{array} \right.
\label{eq:rho_onlygravity}
\end{equation}

\noindent where $x=r/r_s$ and $r_s$ is the scale radius. We sharply truncate the
profile at $r=r_{200}$, so that $M_{TOT} \equiv M_{200}$, where $M_{\rm TOT}$ is the integral
of equation \ref{eq:rho_onlygravity} to infinity, $r_{200}$ is the radius which enclose a mass $M_{200}$ where the density
is 200 times the critical density
$\rho_{c}(z) \equiv 3H(z)^2 /8\pi G$, $H(z)$ is the Hubble function and $G$ the gravitational constant.
The truncation of the initial profile imply a null displacement of the particles beyond $r_{200}$,
 hence we can consider just the particles inside $r_{200}$. Furthermore,
we can avoid the modelling of the background as in \cite{S&T2015},
or the computationally expensive measurement of the 2-halo term e.g. \cite{S&T2018}. \\
We can compute now the four different components of
the final profile. The different density profiles are normalised
such as that for each component $M_i(r)=\int_{0}^{\infty} 4 \pi r^2 \rho_i(r)  dr=f_i \cdot M_{200} $,
so that $\sum  M_i(r)=M_{200}$, thus obviously $\sum f_i=1$. Furthermore we truncate all the
density profile at $r_{200}$, except for the ejected gas.\\
The central galaxy is modelled with a power-law with an exponential cut-off \citep{Mohammed2014},
\begin{equation}
\rho_{\rm CG}(r) = f_{\rm CG} \cdot \frac{M_{200}}{4 \pi ^ {3/2}  R_h  r^2} \cdot \exp \left(- \left(\frac{r}{2R_h} \right)^2 \right),
\label{eq:rho_galaxy}
\end{equation}
where the half-mass radius is approximated as $R_h\approx0.015\cdot r_{200}$ \citep{Kravtsov2018}.
There is no need to truncate the galaxy profile, because the exponential cut-off assures
that the density at $r_{200}$ is practically zero.     \\
The hot, bound gas is modelled assuming hydrostatic equilibrium \citep{Martizzi2013} up to $r<r_{200}/\sqrt{5}$,
after which the gas is considered collisionless, thus following the NFW profile (Eq.~\ref{eq:rho_onlygravity}):

\begin{equation}
\rho_{\rm BG}(x) = f_{\rm BG} \cdot  \left\{ \begin{array}{ll}
      y_0 \cdot \left( x^{-1} \ln(1+x) \right) ^{\Gamma_{eff}(c)} & \hspace{5mm} r<r_{200}/\sqrt{5} \\
      y_1 \cdot x^{-1} (1+x)^{-2} & \hspace{5mm} r \le r_{200}  \\
      0 & \hspace{5mm}  r \ge r_{200} ,\\
      \end{array} \right.
\label{eq:rho_hotgas}
\end{equation}
The effective polytropic index $\Gamma_{\rm eff}$ is defined such that the
hydrostatic gas has the same slope of the NFW at $r=r_{200}/\sqrt{5}$:
\begin{equation}
\Gamma_{\rm eff}(c) = \frac{ (1+3c/\sqrt{5})  \ln (1+c/\sqrt{5})}{(1+c/\sqrt{5}) ln(1+c/\sqrt{5})-c/\sqrt{5}},
\label{eq:gamma_eff}
\end{equation}
where $c=r_{200}/r_s$ is the halo concentration. The normalisation factors $y_0$ and $y_1$
are defined such that the profile is continuous and  $\int_{0}^{\infty} 4 \pi r^2 \rho_{\rm BG}(r)  dr = f_{\rm BG} \cdot M_{200} = M_{\rm BG}(r)$. \\
The ejected gas profile is computed assuming a Maxwell-Boltzmann velocity distribution of
the particles expelled by the AGN, and behave as a constant with an exponential cut-off,
\begin{equation}
\rho_{\rm EG}(r) = \frac{M_{200}}{ (2 \pi r_{\rm ej}^2)^ {3/2} } \exp \left(- \frac{1}{2} \left(\frac{r}{r_{\rm ej}} \right )^2 \right),
\label{eq:rho_ejectedgas}
\end{equation}
where the {\it ejected radius} $r_{\rm ej}$ is the maximum radius reached by the expelled gas:
\begin{equation}
r_{\rm ej} \equiv \eta \cdot 0.75 \, r_{\rm esc},
\label{eq:modelA}
\end{equation}
with $\eta$ as a free parameter. The halo escape radius is estimated assuming
a constant halo escape velocity and a time-scale of a half Hubble time:
\begin{equation}
r_{\rm esc} \equiv \Delta t \cdot v_{\rm esc} \approx \Delta t \sqrt{\frac{8}{3} \pi G \Delta_{200} \rho_{\rm crit}} \approx \frac{1}{2} \sqrt{\Delta_{200}} r_{200}.
\label{eq:escape_radius}
\end{equation}

Finally, we compute the dark matter profile, defined as a piecewise function: a NFW up to a
scale $r'$, a constant $\rho_{DM}(r_{200})$, between $r'$ and $r_{200}$ and 0 afterward. We make sure that the density profile is
continuous at all scales by forcing the matching of the profiles at $r_{200}$ and at $r'$.
Within $r_{200}$, the density is given by the sum of all our BCM components, whereas beyond
$r_{200}$ the only contribution is given by the ejected gas, that must be be summed to the initial
GrO density of the halo: $\sum \rho_{i}(r_{200}) = \rho_{EG}(r_{200})+\rho_{GrO}(r_{200})$.
Since $\rho_{CG}(r_{200})\approx0$, we obtain
$\rho_{DM}(r_{200}) = \rho_{GrO}(r_{200})-\rho_{BG}(r_{200})$;
we force the new NFW to pass through the point $(r',~ \rho_{DM,200})$,
\begin{equation}
\rho_{\rm NFW}(x',\rho_0') = \frac{\rho_0'}{x' ( 1+x')^2},
\label{eq:new_rho_nfw}
\end{equation}
where $x'=r'/r_s$. We also impose the mass conservation, i.e.
\begin{equation}
\int_{0}^{r'} 4 \pi r^2 \rho_{\rm NFW}(r,\rho_0') {\rm dr} + \int_{r'}^{r_{200}} 4 \pi r^2  \rho_{\rm NFW}(r') dr = M_{DM}.
\label{eq:mass_conservation}
\end{equation}

We can obtain $r'$ and the new normalisation $\rho_0'$ minimising the function

\begin{multline}
r'  (r_s + r')^2 \left[ \rho_{\rm GrO}(r_{200}) - \rho_{\rm BG}(r_{200}) \right] \left[ \ln \left( 1+\frac{r'}{r_s}  \right) - \frac{r'}{r'+r_s} \right] = \\
= - \frac{\rho_{\rm GrO}(r_{200}) - \rho_{\rm BG}(r_{200})}{3}  \left( r_{200}^3 - r'^3 \right) +\frac{f_{\rm DM} M_{200}}{4\pi}.
\label{eq:rprime}
\end{multline}

We compute then the expected baryonic back-reaction on the dark matter. We allow
the dark matter profile to relax, so that the particles will expand (contract) depending if the total
gravitational potential is shallower (deeper). Let us call $M_i$ the initial mass contained in a sphere of radius $r_i$,
whereas $M_f$ will be the mass after relaxation inside a final radius $r_f$.
Numerical simulations show that the halo relaxation is not perfectly
adiabatic, i.e. $r_f/r_i \neq M_i/M_f$, but it follows the equation

\begin{equation}
\frac{r_f}{r_i} = 1+a \left[ \left( \frac{M_i}{M_f} \right) ^n -1 \right],
\label{eq:relaxation}
\end{equation}

\noindent where $a=0.3$ and $n=2$ \citep{Abadi2010}. Considering

\begin{equation}
 \left\{ \begin{array}{l}
              M_i (r_i) = M_{\rm OG}(r_i) \\
          M_f (r_f) = f_{\rm DM}M_i+M_{\rm EG}(r_f)+M_{\rm BG}(r_f)+M_{\rm CG}(r_f),\\
                \end{array} \right.
\label{eq:MiMf}
\end{equation}

\noindent we can solve the system of Eq.s ~\ref{eq:MiMf} and ~\ref{eq:relaxation} iteratively
for $\xi=r/r_i$, furthermore imposing that $\xi(r_{200})=1$. The relaxed dark matter mass profile is thus

\begin{equation}
M_{\rm RDM}(r) = f_{\rm DM} \cdot M_{\rm GrO}(r/\xi),
\label{eq:mass_rdm}
\end{equation}

\noindent and the density profile will be

\begin{equation}
\rho_{\rm RDM}(r) = \frac{1}{4\pi r^2 } \frac{d}{dr}M_{\rm RDM}(r).
\label{eq:rho_rdm}
\end{equation}

We now report the components mass fractions used in this work. The dark matter fraction is fixed at the cosmic density value, $f_{\rm RDM} = 1-\Omega_b/\Omega_m$. The central galaxy fraction is given by the parametrisation from abundance matching
 by \cite{Behroozi2013}:

\begin{equation}
 f_{\rm CG} (M_{200}) = \epsilon \left( \frac{M_1}{M_{200}} \right) 10^{g(\log_{10}(M_{200}/M_1)) - g(0)},
\label{eq:f_cg}
\end{equation}

 \begin{equation}
    g(x)= -\log_{10} (10^{\alpha x} +1) + \delta \frac{ (\log_{10} (1+\exp(x)))^\gamma}{1+\exp(10^{-x})}.
 \end{equation}

\noindent We use the best-fitting parameters at $z=0$ given by \cite{Kravtsov2018}, while assuming the redshift dependence given by \cite{Behroozi2013}:

\begin{multline}
\nu(a) = \exp(-4a^2) \\
\log_{10}(M_1) = M_{1,0} +(M_{1,a}(a-1)+ M_{1,z}z ) \nu \\
\log_{10}(\epsilon) = \epsilon_{0} +(\epsilon_{a}(a-1)) \nu + \epsilon_{a,2}(a-1) \\
\alpha = \alpha_0+(\alpha_a(a-1))\nu \\
\delta = \delta_0+(\delta_a(a-1) + \delta_z z)\nu \\
\gamma = \gamma_0+(\gamma_a(a-1) + \gamma_z z)\nu , \\
\end{multline}

\noindent with $M_{1,a} = -1.793$, $M_{1,z}=-0.251$, $\epsilon_{0}=\log_{10(0.023)}$, $\epsilon_{a}=-0.006$, $\epsilon_{a,2}=-0.119$,
$\alpha_0 = -1.779$, $\alpha_a=0.731$, $\delta_0=4.394$, $\delta_a=2.608$, $\delta_z=-0.043$,
$\gamma_0 = 0.547$, $\gamma_a=1.319$, $\gamma_z=0.279$.

We set the characteristic halo mass, $M_{1,0}$, for which the galaxy-to-halo mass fraction is $\epsilon_0$,
as a free parameter of the model. Notice that, however, we quote throughout the paper the derived parameter $M_1$.

The hot gas mass fraction reads

\begin{equation}
f_{\rm BG} (M_{200}) = \frac{\Omega_b/\Omega_m-f_{\rm CG}}{(1+(M_c/M_{200})^\beta)},
\label{eq:f_bg}
\end{equation}

\noindent with $M_c$ and $\beta$ free parameters. Notice that, for $M_c=M_{200}$ we obtain $f_{\rm BG}=0.5 \, \Omega_b/\Omega_m-f_{\rm CG}$, i.e. half of the gas is retained in the halo.

The ejected gas mass fraction is simply
\begin{equation}
f_{\rm EG} (M_{200}) = \Omega_b/\Omega_m-f_{\rm CG}(M_{200})-f_{\rm BG}(M_{200}).
\label{eq:f_eg}
\end{equation}
In this way, high values of $M_c$ imply that all the gas is expelled even from massive haloes,
and vice versa for low values of $M_c$ progressively smaller halos are gas rich. \\
To recap, in our modelling we have four free parameters: $\eta$, directly proportional to
the halo ejected radius, escape radius and critical radius; $M_c$ is the characteristic halo mass
for which half of the gas is retained; $\beta$ describes how fast the depletion of gas
increase going toward smaller haloes; $M_1$, which is the characteristic halo mass that host
a central galaxy of a given mass.

\section{Subsampling of components particles}
\label{app:subsampling}

Once we have the galaxy, hot gas, ejected gas and dark matter analytical profiles,
we can tag and rescale the mass of the gravity-only particles to match those profiles.
In this way, we can get a ``baryonic'' simulation. The main steps of the algorithm are the following:

\begin{itemize}
 \item Compute the theoretical cumulative and differential mass profiles of the different components;
 \item Compute the theoretical bin/total mass fraction for each component $f$;
 \item Count the number of halo particles per radial bin $C$;
 \item The quantity $\mathcal{N} = f \cdot C$ gives the number of particles per radial bin per component;
 \item Iteratively add to $\mathcal{N}$ the particles missing for discretisation (roundoff);
\item The mass of the particles per component and per radial bin is given by
\begin{equation}
m = m_{p} \cdot \frac{C} {\mathcal{N}} \frac{M_c}{M_{bcm}},
\end{equation}

\noindent where $m_{p}$ is the particle mass in the gravity-only original simulation and $M_c$
and $M_{bcm}$ are the component and the total
{\it baryonic correction} differential mass profiles, respectively.

\end{itemize}

\begin{figure}
  \includegraphics[width=.9\linewidth]{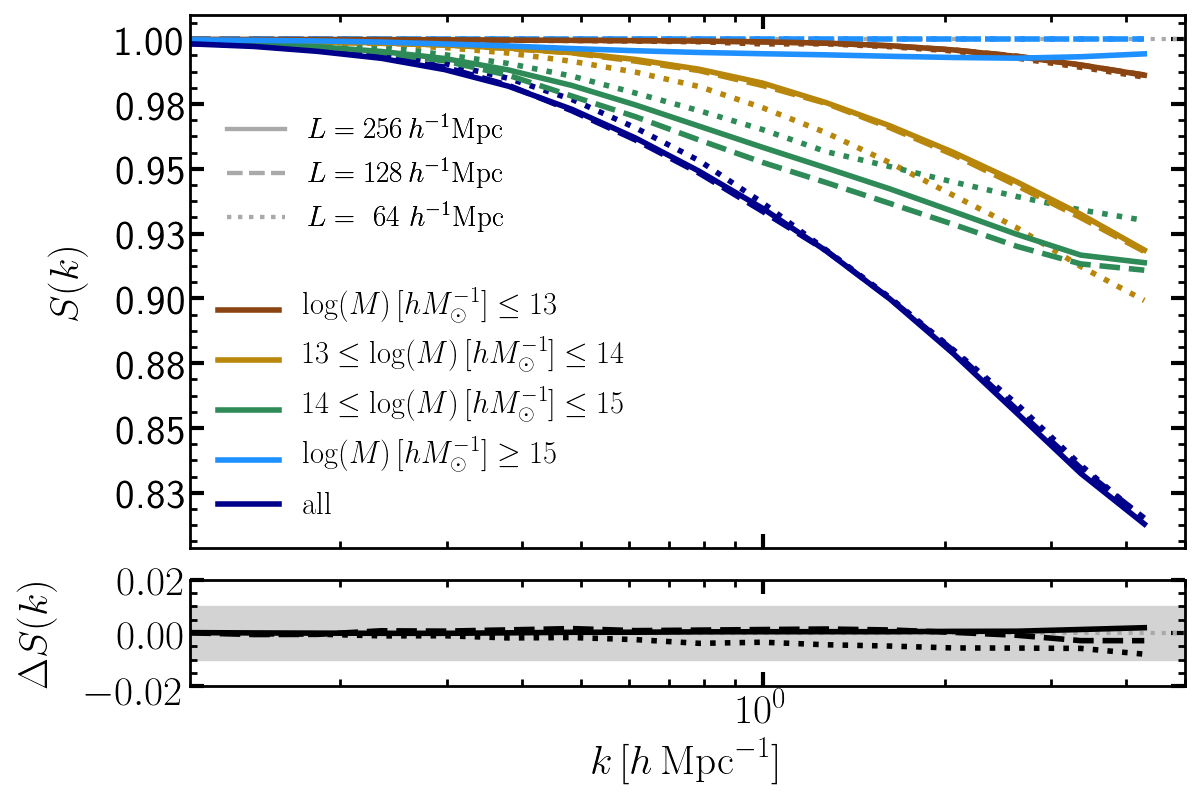}
  \caption{ {\it Upper panel:} Baryon suppression of the matter power spectrum, defined as $S(k) \equiv P_{\rm BCM}/P_{\rm GrO}$ at $z=0$. Solid, dashed and dotted lines
          are computed with simulations of box side  $256$, $128$ and $64 \, \hMpc$ and 768$^3$, 384$^3$ and 192$^3$ particles, respectively.
         Colors are referred to different halo mass bins whit which the baryon corrections
          have been computed, according to the legend.
          {\it Lower panel:} Difference in suppression between two {\it paired and fixed} simulations and
                  a single one, for the three different volumes specified in the legend of the upper panel. }
          \label{fig:convergence}
\end{figure}

\section{Convergence of the {\it baryon} simulation}
\label{app:convergence}

We have test the convergence of the {\it baryonic} suppression by running three simulations with
increasing volume and number of particles, while keeping same resolution. In particular, we have run
simulations with $64 \, \hMpc$, $128 \, \hMpc$ and $256 \, \hMpc$ of box side, with $N=192$, $N=384$
and $N=768$ cubic particles, respectively. All the simulations share the same initial conditions,
and for each different volume we have run two simulations with fixed amplitude and shifted phases
as reported in \S\ref{sec:sim}. In Fig.~\ref{fig:convergence} we show the suppression $S(k)$, defined
as the ratio between baryonic and gravity only matter power spectra, for the three simulations at $z=0$.
We consider a BCM with the following parameters: $M_c=1.2 \cdot 10^{14} ~ \Msun$, $\eta=0.5$, $\beta=0.6$, $M_1=2.2 \cdot 10^{11} \Msun$. We note that even if the cosmic variance is consistently different, this contribution is canceled out at first order
in the ratio of the power spectra, which are consistent with each other well within 1\%. In the bottom panel
of Fig.~\ref{fig:convergence}, we show the difference in suppression between using standard and {\it paired and fixed} simulations.
For the biggest volume considered the suppression is practically the same (solid line),
and even for the smallest volume the difference is well within 1\% (dotted line).

We have investigated moreover the contribution of different halo masses to the total baryonic
suppression. As shown in the upper panel of \ref{fig:convergence}, the biggest relative contribution
is given by haloes of mass $M=10^{14}-10^{15} \, \Msun$ and  $10^{13}-10^{14} \, \Msun$. It appears that
the number of very massive haloes ($M\ge10^{15} \, \Msun$) is not sufficient to produce an effect lager than
1\%, even for the largest simulation considered in this work. Haloes with masses $M\le10^{13} \, \Msun$ have an impact of $\approx$ 2\% on small scales, $k\approx5 \, \ihMpc$.

\bsp	
\label{lastpage}
\end{document}